\documentclass[hyper,letterpaper,11pt]{JHEP3}
\usepackage{amsmath,amssymb,multirow,array}
\usepackage{cite}

\newcommand\beq{\begin{equation}}
\newcommand\eeq{\end{equation}}
\newcommand\be{\begin{equation}}
\newcommand\ee{\end{equation}}

\title{Entanglement Entropy for Probe Branes}

\preprint{\today}

\author{Han-Chih Chang and Andreas Karch \\
Department of Physics, University of Washington, Seattle, WA 98195, USA \\
Email: hanchih@uw.edu, akarch@uw.edu
}

\abstract{
We give a prescription for calculating the entanglement entropy in holographic probe brane systems by systematically taking the leading order backreaction of the probe brane into account. We find a simple compact double integral formula, which is insensitive to many details of the backreaction, most notably the internal space or the non-metric fields sourced by the probe. We validate our method by comparing to exact results in solvable toy models. We also determine the entanglement entropies for a sphere and a strip in the top-down D3/D7 and D3/D5 system. For the sphere the entanglement entropy has also been obtained by other methods and we find perfect agreement.
}

\begin{document}
\tableofcontents

%%%%%%%%%%%%%%%%%
\section{Introduction}
\label{sec:introduction}

Entanglement entropy (EE) has emerged as a very powerful theoretical tool in the studies of topological phases of matter, strongly correlated systems in general and even the quantum nature of gravity. A large class of strongly coupled field theories in which EE can be calculated reliably is provided by holography\cite{Witten:1998qj,Maldacena:1997re,Gubser:1998bc}. Ryu and Takayanagi (RT) introduced in \cite{Ryu:2006bv} a very simple prescription for how to calculate the EE in field theories with a holographic dual. To specify the EE in a field theory with $d-1$ spatial dimensions, one needs to pick a $d-2$ dimensional entangling surface, $\Sigma$, separating (at a given time $t$) the degrees of freedom of the field theory into two subsystems A and B. By tracing over the degrees of freedom in B, one obtains a reduced density matrix for the degrees of freedom in A and vice versa. Even when describing a pure state, the reduced density matrices are mixed due to the loss of information inherent in tracing over a subspace. The standard von Neumann entropy associated with the reduced density matrix is the entanglement entropy. For a zero temperature state, the entanglement entropies associated with subsystems A and B respectively are identical. This entanglement entropy provides a measure of the entanglement present in the original state of the full system. The RT proposal asserts that, in the holographic dual description, the EE is given by
\be
\label{RT}
S_A =\underset{
\{\gamma_A|_{\scriptscriptstyle \partial \gamma_A= \Sigma\}}}{\mbox{Min}}
\frac{\mbox{Area}(\gamma_A)}{4 G_N},
\ee
%\be
%S_A =\min_{ \{ \gamma_A |\partial \gamma_A =\Sigma\} }
%\frac{\mbox{Area}(\gamma_A)}{4 G_N}
%\ee
where $G_N$ is Newton's constant and the $\gamma_A$, whose area determines the EE, is a minimal area surface in the holographic bulk, terminating on the prescribed entangling surface $\Sigma$ on the boundary. In the field theory, one important contribution to the EE is the short range entanglement of the degrees of freedom in the vicinity of the entangling surface. This contribution is sensitive to the details of the short distance physics and is proportional to the area of the entangling surface. In addition to this UV divergent area term, there are several subleading terms, some of which carry universal information about the long range entanglement in the state. Especially in the case of conformal theories, the structure of these terms has been clarified by the holographic calculations. For a recent review on these developments see \cite{Takayanagi:2012kg}.

In this work we give a derivation of the holographic EE for a large class of holographic theories for which so far application of the RT formula has been mostly unsuccessful: probe brane systems \cite{Karch:2002sh,Karch:2000gx}. Probe brane setups describe strongly coupled field theories with special properties. Not only does one need to take a ``large $N$" limit which guarantees a classical dual, one also needs two sectors which scale as different positive powers of $N$. One common class of examples are large $N$ gauge theories with order $N^2$ gluon degrees of freedom coupled to fundamental representation quarks with order $N$ degrees of freedom. In this case, the quarks are still classical, but they act as probes of the glue background: their dynamics adjusts itself to the strongly interacting background provided by the glue, but does not backreact on it.

Probe brane systems have been studied in many different contexts. In applications of holography to nuclear physics, the most successful holographic cousin of QCD, the Sakai-Sugimoto model \cite{Sakai:2004cn}, is based on a probe brane system. For applications to condensed matter physics, probe brane setups can realize a variety of interesting situations. The glue degrees of freedom can act as a heat-bath for the quarks, giving the simplest realization of a  model which allows for dissipation and hence a finite DC conductivity \cite{Karch:2007pd}; the properties of such a system can be engineered to be in qualitative agreement with that of high-T$_c$ superconductors in the strange metal phase \cite{Hartnoll:2009ns}. Probe branes give straight forward realizations of holographic lattices \cite{Kachru:2009xf}, including holographic realizations of a Kondo-lattice, giving a controlled field theory example of a non-Fermi liquid \cite{Jensen:2011su}. Non-Landau phase transitions with an exponential scaling of the order parameter close to the critical point are also easy to realize via probes \cite{Jensen:2010ga}. Probe brane systems also can realize novel phases of compressible matter with peculiar properties: they can display a zero sound pole characteristic of Fermi liquids despite an unusual temperature dependence of the heat capacity \cite{Karch:2008fa,Kulaxizi:2008jx,Karch:2009eb}; they can display the appearance of a moduli space despite the absence of any supersymmetry \cite{Chang:2012ek,Ammon:2012mu}; maybe most interestingly in the current context, they can realize non-relativistic critical points with hyperscaling violating exponent $\theta=d-2$ \cite{Ammon:2012je}. This particular value has been argued in \cite{Ogawa:2011bz,Huijse:2011ef} to be associated with a logarithmic enhancement of the area law for the EE. It would be extremely interesting to see if the well-understood field theory provided by the probe brane system of \cite{Ammon:2012je} bears out this expectation. Last but not least, probe brane systems have been instrumental in giving a holographic realization of many non-trivial topological phases, such as the quantum Hall effect \cite{Fujita:2009kw} and fractional topological insulators \cite{HoyosBadajoz:2010ac}. Non-trivial entanglement is a hallmark of topological states, so calculating the EE in these systems should be a worthwhile exercise.

In principle the RT formula can immediately be applied to probe branes. Even though to leading order in a large $N$ expansion the effect of the stress-energy carried by the probe can be completely neglected, one can systematically incorporate the backreaction of the probe brane on the background geometry by solving Einstein's equation in a $1/N$ expansion with the probe-source included. To get the contribution of the probe degrees of freedom to the EE, one simply needs to re-solve the minimal area problem in this fully backreacted metric and then directly apply the RT prescription eq.\eqref{RT}. For simple toy models \cite{Azeyanagi:2007qj} and highly supersymmetric cases with "fat branes" realized as a scalar lump \cite{Chiodaroli:2010ur}, this procedure has been carried out explicitly for the fully backreacted metric. In this work, we are going to derive a simple compact formula for the leading order contribution of the probe to the EE following this general strategy. Note that our formula therefore directly follows from the RT prescription and is not a separate conjecture. We write our final answer for the EE as a double integral involving a gravitational Green's function, eq.\eqref{doubleintegral}. This is the main result of this work. We validate our integral expression by comparing to two solvable toy models. In the process we have to understand the UV divergences in our integral and describe their physical origin and meaning. We can also immediately deduce from our integral expression that the expressions for the EE in our two bottom-up toy models in fact also applies to two of the most studied probe brane systems in type IIB supergravity despite the complications associated with the internal space.

The organization of the paper is as follows. In Section 2 we review the basic definition of a probe brane system and derive our basic formula. In Section 3 we introduce two simple solvable toy models in which the backreaction can easily be taken into account, which we use later to validate our results. In Section 4 we show that, due to the special properties of the entangling surface, in many cases the deformation of the internal space due to the backreaction can be neglected. This allows us to map several top-down probe brane systems, with known field theory dual, to the results obtained in the two solvable toy models. In Section 5 we apply our formalism to the toy models of Section 3 and hence, by the results of Section 4, also to two well-studied top-down models. We find perfect agreements with the fully backreacted answer for the toy models. In the special case of a spherical entangling surface we also find perfect agreement with results obtained by Jensen and O'Bannon using an alternative method based on the Casini-Huerta-Myers trick.

%%%%%%%%%%%%%%%%%
\section{The Entanglement Entropy of probe branes}
\label{sec:doubleintegral}

\subsection{General Probe Brane Systems}

The generic probe brane setup has a bulk action which includes a Einstein-Hilbert term coupled to a matter Lagrangian\footnote{In here and in the following we denote the dimension of the bulk spacetime as $d+1$. The dual field theory has $d$ spacetime dimensions. If the bulk gravity solution also has an internal factor we refer to spacetime as the lower dimensional space one obtains after compactification. The worldvolume of the probe brane has $n+1$ spacetime dimensions, where clearly $n\leq d$. For the bulk we use coordinates $x^{\mu}$ with $\mu=0,\ldots,d$, and for the worldvolume
of the probe brane $z^{i}$ with $i=0,\ldots,n$. Last but not least, the entangling surface in the field theory has $d-2$ spatial dimensions and is completely localized in time. In the holographic dual the EE associated with this entangling surface is dual to a $d-1$ dimensional extremal area for which we use coordinates $w^a$ with $a=0,\ldots,d-2$. As we will be mostly interested in static geometries we reserve the superscript $^0$ for the radial coordinate, not time, as is common in the literature on gravitational propagators on Anti de-Sitter (AdS). Time is the coordinate with the largest label ($x^d$ and $z^n$; $w$ only runs over spatial coordinates). We'll denote by $\vec{x}^2$, $\vec{w}^2$ and $\vec{z}^2$ the contractions of the non-zero indices with a Kronecker delta (when working in Euclidean signature) or an $\eta$ tensor (when working in Lorentzian signature).},

\beq
S_{bulk}=\frac{1}{16 \pi G_N} \int d^{d+1} x \sqrt{-g} \, \left (R + {\cal L}_{bulk} \right ),
\eeq
and a probe brane action, which typically starts with a tension term, that is a uniform energy per unit volume
\be S_{probe} = T_0 \int d^{n+1} z \, \sqrt{-g_I} {\cal L}_{probe} =T_0 \int d^{n+1}z \, \sqrt{-g_I} \left (1 + \ldots \right ). \ee
Here $\sqrt{g_I}$ denotes is the induced metric on the $n+1$ dimensional worldvolume. The matter in the bulk action should allow for a spacetime with a holographic interpretation characterized by a curvature radius $L$. The simplest example is a pure negative cosmological constant giving rise to an AdS$_{d+1}$ vacuum solution. In this case the dual conformal field theory (CFT) however is not known explicitly. For examples, where the dual field theory is known from the embedding of the duality in string theory, the bulk matter sector is typically more complicated: the gravitational AdS$_5$ $\times$ $S^5$ background dual to ${\cal N}=4$ SYM is accompanied by a constant 5-form flux field strength. The brane action can be almost arbitrary. For D-brane probes one typically has, in addition to the tension term, a Maxwell term for a worldvolume gauge field dual to the conserved particle number on the field theory side. The action in this case contains higher powers of the field strength as well, which are known to sum up into the form of Dirac-Born-Infeld (DBI) action. The important property of a probe brane setup is the following hierarchy of scales:
\be
\frac{L^{d-1}}{G_N} \gg T_0 L^{n+1} \gg 1
\ee
$L/G_N^{1/(d-1)}$ is the curvature radius in Planck units. This quantity appears as an overall prefactor of the bulk action. It being large allows us to approximate quantum gravity in the bulk by the semi-classical saddle point, that is by the solution to the classical equations of motion. Similarly $T_0 L^{n+1}$ appears as the overall prefactor of the probe brane action. It being large ensures that the probe brane action can be treated classically as well. Last but not least, the combination $G_N T_0 L^{n-d+2}$ controls the strength with which the brane stress tensor appears as a source on the right hand side of Einstein's equations (and similarly it controls how much other bulk fields are sourced by the brane). It therefore sets the size of the backreaction of the brane on the background geometry. To leading order, the brane simply is a probe that minimizes its own worldvolume action in a fixed background geometry. The backreaction can systematically be calculated in an expansion in the small dimensionless parameter
\be
\label{t0}
t_0 \equiv 16 \pi G_N T_0 L^{n-d+2}.
\ee
In the field theory, probe brane systems describe setups where we have two different classical sectors (that is sectors which a large number of degrees of freedom) with a hierarchy between them. One large class of examples including \cite{Karch:2000gx,Karch:2002sh,Sakai:2004cn} coupling fundamental matter to a large $N$ gauge theory. In this case typically $L^{d-1} G_N \sim N^2$, as there are order $N^2$ glue degrees of freedom, and $T_0 L^{n+1} \sim N$, as there are of order $N$ degrees of freedom in the fundamental matter. The order $N^2$ and $N$ pieces in physical quantities like the free energy or, of interest here, the EE are determined by classical physics. These ``flavor" probe branes are not the only examples that display such a hierarchy of scales. Another important example is the fundamental F1 string: here $n=1$ and $T_0 L^2 \sim \sqrt{\lambda}$ where $\lambda$ is the 't Hooft coupling of the dual gauge theory, which also needs to be taken large in the large $N$ limit for a good supergravity description to exist in that case.

\subsection{Calculating the EE for probe branes}

In principle, the EE for Einstein gravity coupled to brane sources directly follows from the RT formula. One needs to calculate the full backreaction of the probe brane, and then re-solve the minimal area problem for the EE in the fully backreacted geometry. In the probe limit this calculation should however simplify dramatically. We only need the leading order backreaction of the brane on the background geometry, and then calculate the change in area of the minimal area defining the EE in the original background due to this small change in the background metric.

In terms of the probe-brane stress tensor
%\be
%T^{\mu \nu}_{probe} = \frac{2}{\sqrt{-g}} \frac{\delta \left ( \sqrt{-g} {\cal L}_{probe} \right ) }{\delta g_{\mu %\nu}} \delta(x^{\mu} - x_P^{\mu}),
%\ee
\be
T^{\mu \nu}_{probe} =
\frac{2}{\sqrt{-g_I}} \frac{\delta \left ( \sqrt{-g_I} {\cal L}_{min} \right ) }{\delta g_{\mu \nu}}
\Bigg|_{x^{\mu} \rightarrow x^{\mu}_P(z^i)},
\ee
where $x_P^{\mu}(z^i)$ describes the embedding of the probe, the backreacted metric to leading order in the
backreaction can be written as
\be
\label{perturbedmetric}
(\delta g)_{\mu \nu} = (8 \pi G_N T_0) \int d^{n+1} z \, \sqrt{g_I} \, G_{\mu \nu \rho \sigma} T^{\rho \sigma}_{probe}, 
\ee
where $G_{\mu \nu \rho \sigma}$ is the appropriate Green's function of linearized Einstein gravity. For static backgrounds, which we will be mostly concerned with, one can use the Euclidean signature geometry and use the unique Green's function that is regular inside the holographic spacetime and obeys Dirichlet boundary condition at the boundary. For pure AdS, this Green's function has been determined in a very convenient form in \cite{D'Hoker:1999jc}; we'll review the construction in detail when discussing specific examples in the following sections. For time-dependent backgrounds we should use the retarded Green's function.

The minimal area is described by an embedding $x_M^{\mu}(w^a)$ which can be derived from an ``action"
\be
\label{action}
S_{min} = \frac{1}{4 G_N} \int d^{d-1} w \, \sqrt{\gamma} \equiv \frac{1}{4 G_N} \int d^{d-1} w \, {\cal L}_{min},
\ee
where $\gamma$ is the determinant of the induced metric. The EE according to the RT formula is now simply the on-shell value of this action. Note that while RT was originally derived for static backgrounds, it has been argued in \cite{Hubeny:2007xt} that in the time-dependent case the EE is still given by an extremal surface and so the action eq.\eqref{action} still applies in this case. The on-shell Lagrangian for the minimal surface depends both on the embedding function $x_M^{\mu}(w^a)$ and the background metric. To calculate the change in the EE due to the backreaction of the brane we can hence write
\be
\delta S_{min} = \frac{1}{4 G_N} \int d^{d-1} w \, \sqrt{\gamma} \, \left (  \frac{T^{\mu \nu}_{min}}{2} (\delta g)_{\mu \nu} + \frac{\delta {\cal L}_{min}}{\delta x_M^{\mu}} \delta x_M^{\mu} \right ).
\ee
Similar to the probe brane we have defined the ``stress tensor"
%\be
%T^{\mu \nu}_{min} = \frac{2}{\sqrt{-g}} \frac{\delta \left ( \sqrt{-g} {\cal L}_{min} \right ) }{\delta g_{\mu \nu}} %\delta(x^{\mu} - x_M^{\mu}).
%\ee
%
\be
T^{\mu \nu}_{min} =
\frac{2}{\sqrt{-\gamma}} \frac{\delta \left ( \sqrt{-\gamma} {\cal L}_{min} \right ) }{\delta g_{\mu \nu}}
\Bigg|_{x^{\mu} \rightarrow x^{\mu}_M(w^a)}.
\ee
For the minimal area surface, the stress tensor is proportional to the variation of the determinant of the induced metric. For the probe brane we expect such a term to be present as well due to the standard tension term, but extra contributions, e.g. due to the worldvolume gauge fields, are also allowed.

$\delta S_{min}$ can be simplified dramatically by noting that $\frac{\delta {\cal L}_{min}}{\delta x_M^{\mu}}$ vanishes when evaluated on the unperturbed minimal area due to the equations of motion\footnote{This fact has also recently been observed in \cite{Nozaki:2013vta,Wong:2013gua} who also studied the response of the EE to small metric perturbations.}. Plugging in our expression eq.\eqref{perturbedmetric} for the perturbed metric we arrive at the following compact expression of the entanglement entropy
\be
\label{doubleintegral}
S_A = (\pi T_0) \int (d^{d-1} w \sqrt{\gamma}) \, (d^{n+1} z  \sqrt{g_I}) \left (  T^{\mu \nu}_{min} G_{\mu \nu \rho \sigma} T^{\rho \sigma}_{probe} \right ) .
\ee
This simple double integral gives the EE for a generic probe brane system. It can be thought of as the gravitational potential energy between the probe brane and the ``energy density" of the minimal area surface. The formula is valid as long as the only source of $\delta g$, to leading order in $t_0$, comes from the stress tensor on the probe brane. If the probe brane sources bulk fields other than the metric, there are additional contributions one needs to worry about. If we denote bulk fields other than the metric (e.g. the $p$ form potentials of type IIB supergravity) collectively by $\Phi_{bulk}$, the backreaction of the flavors will induce $\Phi_{bulk} \sim t_0$ together with $\delta g \sim t_0$ for all the $\Phi_{bulk}$ that are sourced by the brane. For example, a D$p$ brane in IIB supergravity will source the dilaton and the $p+1$ form Ramond-Ramond gauge field
it is charged under. If the bulk stress-energy tensor has a term linear in $\Phi_{bulk}$, this order $t_0$ change in $\Phi_{bulk}$ drives a change in the metric that is also of order $t_0$ and would have to be included. Fortunately, the Einstein-frame stress tensor for, say, type IIB supergravity is quadratic and higher order in fields, so the stress tensor is at least quadratic in $\Phi_{bulk}$ and so this ``secondary" effect on the metric due to the other brane sources is negligible. The only exception to this statement arises if we are studying a background in which $\Phi_{bulk}$ is turned on before we add the probe brane. In this case, even though the stress tensor is quadratic in the full $\Phi_{bulk}$, there is a term linear in the probe-sourced $\delta \Phi_{bulk}$ when combined with a background $\Phi_{bulk}$. As an example, in the $AdS_5 \times S^5$ solution dual to ${\cal N}=4$ supergravity, the background has a non-trivial 4-form gauge field $C_4$ turned on. Correspondingly, our formula does not apply to branes that source $C_4$. For D7 branes and D5 branes without worldvolume gauge fields our formula is applicable; however, for D3 branes, D5 brane with non-vanishing $F$ and D7 branes with non-vanishing $F\wedge F$, potentially it is not\footnote{The stress tensor associated with $C_4$, written in terms of $H_5 = d C_4$, has terms of the structure $H_{\mu \ldots} H_{\nu}^{\ldots}$ and of the structure $g_{\mu \nu} H^2$. Since the background metric is diagonal, the $C_4$ sourced by the brane needs to share at least 3 indices with the background $C_4$ to have a chance to contribute a non-trivial term to the stress tensor at order $t_0$. For example, the interesting case of a D5 with a field strength $F \propto dr \wedge dt$ turned on along the worldvolume (corresponding to the finite density holographic quantum liquid of \cite{Karch:2008fa}) sources $C_4$ with two legs along the AdS direction and two legs in the internal space. Since the background $C_4$ only has components either entirely in the internal space or entirely in AdS, this particular worldvolume field strength does not give an order $t_0$ term in the stress tensor and so is compatible with our formula even though $C_4$ is sourced.}. Another case in which our analysis does not apply due to secondary backreaction is a D6 brane probe \cite{Hikida:2009tp,Gaiotto:2009tk,Hohenegger:2009as,Ammon:2009wc,Jensen:2010vx} in ABJM \cite{Aharony:2008ug} (the D6 sources the 2-form RR field strength, which in ABJM has a non-trivial background).

The Green's function in the expression above falls off sufficiently fast near the boundary to ensure that the above integral is finite for sources that do not extend out to the boundary. However, most examples of probe branes of interest {\it do} involve probe branes extending all the way to the boundary. This is the case whenever the probe brane describes the addition of matter to the boundary field theory, such as in the D3/D5 system \cite{Karch:2000gx}, the D3/D7 system \cite{Karch:2002sh} or the Sakai-Sugimoto model \cite{Sakai:2004cn}. Probe branes completely localized in the bulk of the holographic spacetime, such as e.g. a single D3 brane probe\footnote{As a D3 brane probe sources $C_4$ in this particular instance our formula wouldn't apply to begin with. See the discussion in the preceding paragraph.} at a fixed radial position in AdS$_5$ $\times$ $S^5$ or the D7 brane dual to the quantum Hall effect of \cite{Fujita:2009kw}, correspond to states of the dual field theory. While in the latter case of a radially localized probe the $z$ integral is UV finite, the more interesting case of a probe extending to the boundary has a divergence in the $z$ integral that needs to be tamed. In addition, the minimal area defining the EE always extends all the way to the boundary. Correspondingly its ``stress tensor" doesn't fall off near the boundary and so the $w$ integral is always UV divergent and needs to be regulated. This UV divergence is physical, capturing the underlying structure of the entanglement. Like the leading divergence of EE itself, the correction to the EE due to the probe brane is sensitive to short distance physics as it is dominated by the short range entanglement of nearby probe degrees of freedom. We'll discuss both these UV sensitivities in more detail below when we look at explicit examples.

%%%%%%%%%%%%%%%%%%%%%%%%%%%
\section{Two simple solvable examples}
\label{sec:toymodels}

There are two simple examples documented in the literature where the full backreaction of a toy model of probe brane can be given in a simple closed-form expression and hence one can easily calculate the full entanglement entropy, not just in the probe limit. We will use the exact answers in these two examples to validate our method. Both are based on the simplest holographic bottom-up action which is not directly obtained from string theory and correspondingly the dual field theory is not explicitly known. The gravitational action in both cases is simply Einstein gravity with a negative cosmological constant in $d+1$ dimensions:
\beq
S=\frac{1}{16 \pi G_N} \int d^{d+1} x \sqrt{g} \, \left (R + \frac{d (d-1)}{L^2} \right ).
\eeq
The vacuum solution for this action is AdS$_{d+1}$ with curvature radius $L$. We write the metric on AdS$_{d+1}$ as
\be
\label{adsmetric}
ds^2=\frac{L^2}{(x^0)^2} \delta_{\mu \nu} dx^{\mu} dx^{\nu},
\ee
where the reader should keep in mind that $x^0$ is the (spatial) radial coordinate. $\eta_{\mu \nu}$ is mostly plus and has -1 as its last diagonal entry. For static configuration we will often be interested in Euclidean AdS in which case $\eta_{\mu \nu}$ is replaced with $\delta_{\mu \nu}$.
The probe brane action only has the standard tension term
\be 
S_{probe} = -T_0 \int d^{n+1}z \, \sqrt{g_I} \label{sprobe}, 
\ee
where $\sqrt{g_I}$, as before, is the induced metric on the $n+1$ dimensional worldvolume. The two special cases we will be considering are $n=d$ (a spacetime filling probe brane) and $n=d-1$ (a codimension-1 probe brane).

\subsection{Spacetime filling probe branes}

The case of a spacetime filling probe ($n=d$) has been studied in detail in \cite{Karch:2008uy} as an exactly solvable model for probe branes. The important point here is that a spacetime filling brane of this type simply corresponds to a shift in the cosmological constant and hence the exact solution is again AdS$_{d+1}$ with a shifted curvature radius $l$ given by \cite{Karch:2008uy}
\be
\label{lshift}
l = L \left ( 1 + \frac{t_0}{2 d (d-1)} \right ).
\ee
This can be rewritten as a first-order shift in the metric
\be
\label{codimzero}
(\delta g)_{\mu \nu} = \frac{t_0 L^2}{d (d-1)} \frac{\delta_{\mu \nu}}{(x^0)^2}.
\ee
For any entangling surface the EE in the full spacetime geometry can be obtained from the one in the original AdS by implementing the simple change eq.\eqref{lshift}.

\subsection{Codimension-1 ``RS" type probe branes}

For $n=d-1$ the simple bottom-up model reduces to the famous Randall-Sundrum setup \cite{Randall:1999ee,Randall:1999vf}. Depending on the tension, the defect worldvolume can be AdS, dS or Minkowski space. In the probe limit we are always automatically in the limit of an ``undercritical" tension, where the small tension of the brane gives a negligible contribution to the induced cosmological constant on the brane and so the worldvolume is AdS. This scenario has been shown to be dual to a conformal field theory with defect or boundary in \cite{Karch:2000gx}. The EE for this model in $d=2$ has been calculated in \cite{Azeyanagi:2007qj}. For all RS (that is codimension-1) setups the fully backreacted spacetime metric can be easily found using the Israel junction equations \cite{Israel:1966rt}. As the sources are delta-function localized, the spacetime can be taken to be a slice of AdS on both sides of the defect. These two spaces will be glued together across the location of the brane. Of course the metric should be continuous across the interface, but it's first derivative however will not be, due to the delta function source. Integrating Einstein equations across the defect, one finds that the jump in the extrinsic curvature is given by the brane stress tensor,
\be
\left . \left ( K_{i j} - g_{i j} K \right ) \right |_{r-\epsilon}^{r+\epsilon}  = 8 \pi G_N T_{i j},
\ee
where $r$ is the coordinate normal to the hypersurface of the brane and $\epsilon$, as usual, is an infinitesimally small positive number. To find a solution to the Israel jump equation it is easiest to write AdS$_{d+1}$ in AdS$_d$ slicing:
\be
\label{adsslicing}
ds^2 = d r^2 + \cosh^2 \left ( \frac{r-c}{L} \right ) \, ds^2_{AdS_d}.
\ee
Here $ds^2_{AdS_d}$ denotes the metric on an AdS$_d$ with curvature radius $L$. For a globally AdS$_{d+1}$
spacetime the constant $c$ can be absorbed by shifting the $r$ coordinate. For the RS setup it is however convenient
to use $c$ to locate the brane hypersurface at $r=0$. The piecewise AdS geometry, that is the fully backreacted solution in response to the brane source, can now be written as
\be
\label{ourmetric}
ds^2 = dr^2 +  \cosh^2 \left ( \frac{|r|-c}{L} \right ) \, ds^2_{AdS_d}.
\ee
Clearly this is locally AdS$_{d+1}$ away from $r=0$. At $r=0$ the jump equation reads
\be 
\frac{2 (d-1)}{L}  \sinh \left( \frac{c}{L} \right ) \,  g^{AdS_d}_{ij} = 8 \pi G_N T_{ij} = 8 \pi G_N T_0 \cosh\left( \frac{c}{L} \right ) g^{AdS_d}_{ij} ,
\ee
where in the last step we used that the matter action is given by eq.\eqref{sprobe}. The Israel jump condition hence gives us $c$ in terms of the tension as
\be
\label{tzero}
t_0 = 4 (d-1) \tanh \left( \frac{c}{L} \right ) \approx \frac{4 (d-1) c}{L}.
\ee
In the last step we used that in the probe limit $t_0$ and hence $c$ are small,  so the hyperbolic tangent function can be approximated by its argument. The first equality in eq.\eqref{tzero} however is exact even away from the probe limit. In the probe limit we can write the change in the metric as
\be
\label{ourmetricperturbed}
\delta g = -  \frac{c}{L} \sinh \left ( 2 \frac{|r|}{L} \right ) \, ds^2_{AdS_d} =
- \frac{t_0}{4 (d-1)} \sinh \left (2 \frac{|r|}{L} \right ) \, ds^2_{AdS_d}.
\ee
To perform calculations in the AdS$_{d+1}$ background in the standard coordinate system used in eq.\eqref{adsmetric}, one can transform $\delta g$ into the $x^{\mu}$ coordinates of eq.\eqref{adsmetric}. Parametrizing the AdS$_{d}$ metric as
\be
ds^2_{AdS_d}= L^2 \left(  \frac{dy^2 + d \vec{x}^2}{y^2} \right),
\ee
where $\vec{x}$ stands for $x^{2}, \ldots , x^d$, we see that the two coordinate systems given in eq.\eqref{adsmetric} and eq.\eqref{adsslicing} (for $c=0$) are related to each other by the following change of coordinates:
\be x^0 = \frac{y}{\cosh(r/L)}, \quad \quad x^1 = y \tanh(r/L). \ee
Applying the same change of coordinates to $\delta g$ allows us to write its components in the $x^{\mu}$ coordinate system as
\be
\label{codimone}
(\delta g) = - \frac{L^2 t_0}{2 (d-1) (x^0)^2} \frac{|x_1|}{\sqrt{(x^0)^2+(x^1)^2}} \left ( d \vec{x}^2 + \frac{(x^1 dx^1 + x^0 dx^0)^2}{(x^0)^2 + (x^1)^2} \right ).
\ee

\section{Internal space}
\label{sec:internal}

In the last section we introduced two toy models with the motivation of providing gravity-plus-probe systems allowing exact determination of the EE to all orders in the backreaction. We need these examples in order to validate our method. Probes with a known string embedding are typically much more complicated. For example, the well-studied D3/D7 \cite{Karch:2002sh} and D3/D5 \cite{Karch:2000gx,DeWolfe:2001pq} systems correspond to codimension-2 and codimension-4 branes respectively wrapping an AdS$_5$ $\times$ $S^3$ or AdS$_4$ $\times$ $S^2$ submanifold in an AdS$_5$ $\times$ $S^5$ background geometry. Their backreaction will induce a change in the full 10 dimensional geometry with non-trivial dependence on the internal $S^5$ coordinates. The fully backreacted solutions have been worked out for the D3/D7 system in \cite{Kirsch:2005uy} and for the D3/D5 system in \cite{D'Hoker:2007xy,D'Hoker:2007xz,Aharony:2011yc}. While available, these fully backreacted solutions are rather complicated and working out solutions to the minimal area problem relevant for the EE in these backgrounds is cumbersome\footnote{
For D5 and D7 probes where the probes are uniformly smeared over the internal directions, 
fully backreacted solutions have been found in \cite{Bigazzi:2005md,Casero:2006pt}. 
Such smearing is possible when there is a large number of flavor branes with 
$N_c \gg N_f \gg 1$ where the probes are uniformly distributed in a spherical
 configuration around the color branes.
}.

As we will explain in more detail here, our general result eq.\eqref{doubleintegral}, for the leading order correction to the EE in the probe expansion, allows us to demonstrate that the answers we obtain in the toy models are, in fact, valid for a large class of branes which are non-trivially embedded into the full 10 (or higher) dimensional space-time. In particular, the EE calculation for the D3/D7 system (to leading order in $t_0$) can be mapped exactly to our codimension-0 toy model, whereas the EE for the D3/D5 system (again to leading order in $t_0$) can be mapped to the codimension-1 toy model. To this order, the fully backreacted solutions of \cite{Kirsch:2005uy,D'Hoker:2007xy,D'Hoker:2007xz,Aharony:2011yc} are not needed.

To prove this assertion\footnote{
This discussion closely parallels the one in \cite{Mateos:2006yd,Bigazzi:2013jqa},
where a similar independence of the internal geometry has been observed 
for other holographic quantities.
}, we need to use the representation of a Green's function as a sum over eigenfunctions. The Green's function $G_{\mu \nu \mu' \nu'}$ of (trace reversed) linearized Einstein gravity obeys a differential equation,
\be
\label{definegreen}
{\cal W}_{\mu \nu}^{\, \, \, \, \, \, \lambda \rho} G_{\lambda \rho \mu' \nu'} =
\left ( g_{\mu \mu'} g_{\nu \nu'} + g_{\mu \nu'} g_{\nu \mu'} - \frac{2}{d-1} g_{\mu \nu} g_{\mu' \nu'} \right)
\delta(x,x') + D_{\mu'} \Lambda_{\mu \nu \nu'} + D_{\nu'} \Lambda_{\mu \nu \mu'},
\ee
where $\delta(x,x')$ is the appropriate curved space Dirac delta function.
Here ${\cal W}_{\mu \nu}^{\, \, \, \, \, \, \lambda \rho}$ is a linear second order differential operator acting on rank-2 tensors $h_{\mu \nu}$. Its detailed form can be found in any standard textbook and will not be important for our argument. In the example of AdS$_{d+1}$ that will be of most interest to us below, we have \cite{D'Hoker:1999jc}
\be
\label{w}
{\cal W}_{\mu \nu}^{\, \, \, \, \, \, \lambda \rho} h_{\lambda \rho} =
- D^{\sigma} D_{\sigma} h_{\mu \nu} - D_{\mu} D_{\nu} h^{\sigma}_{\sigma} +
D_{\mu} D^{\sigma} h_{\sigma \nu} + D_{\nu} D^{\sigma} h_{\mu \sigma} - 2 (h_{\mu \nu} -
g_{\mu \nu} h^{\sigma}_{\sigma}).
\ee
The pure diffeomorphism terms on the right hand side of eq.\eqref{definegreen}, represented by $\Lambda_{\mu \nu \mu'}$, can be eliminated by a coordinate transformation on the primed coordinates. With appropriate boundary conditions ${\cal W}_{\mu \nu}^{\, \, \, \, \, \, \lambda \rho}$ is a hermitian operator and so its eigenfunctions form a complete orthonormal basis\footnote{This is certainly true
for the Euclidean operator, which is appropriate for static calculations as we perform here. For time-dependent backgrounds one typically imposes purely in-falling boundary conditions \cite{Son:2002sd} at the horizon ruining the Hermiticity of ${\cal W}_{\mu \nu}^{\, \, \, \, \, \, \lambda \rho}$. Correspondingly, our arguments here may not be applicable to the retarded Green's function in that case.}.

For product manifolds these eigenfunctions factorize mode by mode. Let us first exhibit the consequences of this factorization for the simpler example of a scalar field. In that case the analog of eq.\eqref{definegreen} reads
\be
 W G(x,x') = \delta(x,x'),
\ee
where $W$ this time is simply the (curved space) Laplacian. For the product manifold we can write
\be
W =W^S + W^I,
\ee
where $W^{S/I}$ only acts on the spacetime/internal part.
One can obtain a representation of $G$ by first solving the eigenvalue problem for W:
\be 
W\psi^S_{m}(x_S) \psi^I_{n} (x_I) = (E_{m} + E_n) \psi^S_{m}(x_S) \psi^I_n(x_I).
\ee
Here $x_{S/I}$ collectively stand for all the coordinates in the spacetime/internal factor of the product manifold, and we already employed a separation of variables ansatz for the eigenfunctions:
\be W^S \psi_m^S(x_S) = E_m \psi_m^S(x_S), \quad \quad  W^I \psi_n^I(x_I) = E_n \psi_n^I(x_I),\ee
where $n$ and $m$ are labeling the eigenmodes. One particular mode that will play an important role in what follows is the zero mode of the internal space. The constant function $\psi^I_0 = V_I^{-1/2}$, where $V_I$ denotes the volume of
the internal manifold, is clearly annihilated by the Laplacian and so is an eigenfunction with eigenvalue
$E_0=0$.

Since the eigenfunctions of the hermitian operator $W$ form a complete orthonormal basis, we can immediately write
\be
\label{grep}
G(x_S,x_I,x'_S,x'_I) = \sum_{n,m} \frac{\psi^S_{m}(x_S) \psi^S_{m}(x'_S) \psi^I_n(x_I) \psi^I_n(x'_I)}{E_{n} +E_{m}}.
\ee
This representation of $G$ is of huge help when integrating against sources which themselves factorize. In particular, if the source $T(x_S,x_I)$ is constant as a function of the internal directions, we can write $T(x_S) = V_I^{1/2}  \psi^I_0(x_I) T(x_S)$. When integrating $T$ against $G$, orthonormality of the modes now picks out the corresponding zero mode from the Green's function:
\be \int_{x_S,x_I} \sqrt{g}\,  G(x_S,x_I,x'_S,x'_I) T(x_S) = \sqrt{V_I} \psi^I_0(x'_I) \, \int_{x_S} \sqrt{g_S} \, G(x_S,x'_S) T(x_S), \ee
where
\be
\label{greps}
G(x_S,x'_S) = \sum_{m} \frac{\psi^S_{m}(x_S) \psi^S_{m}(x'_S) }{E_{m}}
\ee
is the Green's function on the spacetime factor.

If we want to calculate the scalar analog of our double integral eq.\eqref{doubleintegral}, we can apply eq.\eqref{greps} as long as one of the two scalar sources (which we call $T_{min}$ and $T_{probe}$ in parallel with  eq.\eqref{doubleintegral}) are constant on the internal space. For concreteness, assume $T_{min}$ is constant. We can now use eq.\eqref{greps} to write (using $\psi^I_0(x'_I)=V_I^{-1/2}$)
\begin{eqnarray} \nonumber
I &\equiv& \int_{x_S,x_I,x'_S,x'_I} \sqrt{g} \sqrt{g'} \, \,  T_{min}(x_S) G(x_S,x_I,x'_S,x'_I) T_{probe}(x'_S,x'_I)= \\ &=&
\int_{x_S,x_S'} \sqrt{g_S} \sqrt{g'_S} \, T_{min}(x_S) G(x_S,x'_S) T_{probe}^{eff}(x'_S),
\end{eqnarray}
with
\be
T_{probe}^{eff}(x'_S) = \int_{x'_I} \, \sqrt{g_{I}} \, T_{probe}(x'_S,x'_I).
\ee
That is, $I$ reduces to the analogous double integral in the lower dimensional spacetime with the lower dimensional propagator coupling $T_{min}$ to an effective source $T_{probe}^{eff}$, which is the integral of $T_{probe}$ over the internal space. If $T_{probe}$ is constant on the internal space as well, the effective source simply picks up a factor of the internal volume.

How much does this structure carry over to the gravitational case of interest for the EE?
The gravitational Green's function can still be written in terms of modes. For example in the early work
on (anti) de-Sitter space \cite{Turyn:1988af,Allen:1986tt} the gravitational Green's function was written as a superposition of eigenmodes of ${\cal W}_{\mu \nu}^{\, \, \, \, \, \, \lambda \rho}$:
\be
\label{decomp}
G_{\mu \nu \rho \sigma} = \sum_{k=0}^{\infty} a_k h^k_{\mu \nu} h^k_{\rho \sigma}+
\sum_{k=1}^{\infty} b_k V^k_{\mu \nu} V^k_{\rho \sigma}+
\sum_{k=2}^{\infty} c_k W^k_{\mu \nu} W^k_{\rho \sigma}+
\sum_{k=0}^{\infty} d_k \chi^k_{\mu \nu} \chi^k_{\rho \sigma}+
\sum_{k=2}^{\infty} e_k [ \chi^k_{\mu \nu} W^k_{\rho \sigma} + \leftrightarrow].
\ee
Here $h$, $V$, $W$, and $\chi$ are all orthonormal eigenmodes of ${\cal W}_{\mu \nu}^{\, \, \, \, \, \, \lambda \rho}$ with different tensor structure: $W$ and $V$ are traceless tensors and correspond to longitudinal and shear tensor modes, $h$ are the genuinely spin-2 transverse traceless modes, and $\chi$ are pure trace modes.

In a product spacetime, a similar decomposition of the propagator into eigenmodes with different tensor structures still exists. In particular, one of the terms in the expression for $G_{\mu \nu \rho \sigma}$ will
come from gravitational fluctuations with both indices on the ``spacetime" part of the product manifold so that the mode is a genuine tensor in spacetime but a scalar in the internal space:
\be
G_{\mu \nu \rho \sigma} (x_S,x_I,x'_S,x'_I) = \sum_{n,m} \frac{\psi_{\mu \nu}^{m,S}(x_S) \psi_{\rho \sigma}^{m,S}(x'_S) \psi^I_n(x_I) \psi^I_n(x'_I)}{E_{n} +E_{m}} + \ldots \,.
\ee
Here the $\psi^I_n$ are the scalar eigenmodes in the internal space. Clearly for this particular contribution to the propagator, the same simplifications as in the scalar analog integral occur, when we integrate against a source $T_{min}^{\mu \nu}$ that only produces a metric perturbation which is constant on the internal space and only has non-vanishing components with both indices in spacetime. Unfortunately a generic source will give rise to non-vanishing components of {\it all} modes, including those which are vectors or tensors on the internal space.
For the EE however one of our sources, $T^{\mu \nu}_{min}$, is very special in that it is the stress tensor of a codimension-2 minimal surface wrapping the entire internal manifold. It being a minimal surface implies that
\be \label{tmn} T^{\mu \nu}_{min} = \alpha_0\mbox{~} \gamma^{ab} X^{\mu}_{,a} X^{\nu}_{,b}, \ee
where $\gamma^{ab}$ still denotes the induced metric, $\alpha_0$ is a constant, and $a$, $b$ label the worldvolume directions.
From eq.\eqref{tmn} we see immediately that, for a codimension-2 minimal surface, $T^{\mu}_{\mu}=(D-2) \alpha_0$, where $D$ stands for the total dimension of the product spacetime.
The fact that the minimal area wraps the internal space (that is, the minimal area itself is of the form ${\cal N} \times I$ where ${\cal N}$ is a minimal submanifold of the spacetime factor whereas $I$ is the entire internal space) implies that all internal components of the stress tensor are just $\alpha_0$ times the spacetime metric; in particular this implies that all mixed spacetime/internal components of $T_{\mu \nu}^{min}$  vanish. Last but not least the fact that the minimal area is codimension-2 also implies that for the trace reversed stress tensor,
\be
\tilde{T}_{\mu \nu} = T_{\mu \nu} - \frac{1}{D-2} g_{\mu \nu} T^{\rho}_{\rho},
\ee
all internal components of $\tilde{T}^{min}_{\mu \nu}$ vanish.

This statement is important, since the trace reversed Einstein equations
\be R_{\mu \nu} = \tilde{T}_{\mu \nu} \ee
then translate into that the stress tensor of the EE minimal surface does not force any change in the Ricci tensor associated with the internal space, and the unperturbed internal metric solves the full equations of motion. None of the non-trivial tensor or vector modes on the internal space are sourced. For general Freund-Rubin compactifications \cite{Freund:1980xh}, which includes the AdS$_5$ $\times$ $S^5$ background of type IIB supergravity and its AdS$_{4/7}$ $\times$ $S^{7/4}$ M-theory cousins, this can e.g.\ be seen explicitly in the work of \cite{DeWolfe:2001nz} where the full fluctuation spectrum in such compactifications was determined. A trace reversed stress tensor, with only components on the AdS part turned on, sources only modes which are scalar spherical harmonics on the internal sphere.

So for the case of interest where one of the stress tensors in the double integral of eq.\eqref{doubleintegral} is $T^{min}_{\mu \nu}$ with the special properties elucidated above, the double integral can be reduced, just as in the scalar case, to the same double integral performed only over the spacetime part, with the probe stress tensor replaced by an effective probe brane stress tensor obtained from integrating
the full probe stress tensor over the internal space
\be
\label{effective}
T_{\mu \nu}^{probe,eff} = \int_{x_I} \sqrt{g_I} \, T_{\mu \nu}^{probe}.
\ee
Here $\mu$, $\nu$ in eq.\eqref{effective} only run over the spacetime indices of the product manifold. The effective probe stress tensor has no internal components. The full probe stress tensor generically has non-trivial internal components (even in trace reversed form), but as the minimal area did not source any metric perturbation in the internal space, the internal components of the probe stress tensor have nothing to couple to, and do not contribute to the double integral.

For probe stress tensors that are proportional to the metric on the internal space (as one would get if the probe is governed by a DBI action with no fluxes on the internal space turned on), the effective probe stress tensor can be derived from a  DBI action solely defined on the spacetime part of the product space, with an effective tension given by the full probe tension times the volume of the internal space. For the D3/D7 and the D3/D5 systems the values of the effective tension (both $T_0$ and $t_0$ defined in eq.\eqref{t0}) are
\begin{center}
\begin{tabular}{|c|c|c|}
  \hline
  probe & $T_0$ & $t_0$ \\
  \hline
  D5 on AdS$_4$ $\times S^2$ & $ N_f N \frac{\sqrt{\lambda}}{2 \pi^3}$&$ \frac{N_f}{N} \frac{4 \sqrt{\lambda}}{ \pi}$ \\
   D7 on AdS$_5$ $\times S^3$  & $ N_f N\frac{\lambda}{(2 \pi)^4}$ & $ \frac{N_f}{N} \frac{\lambda}{2 \pi^2}$\\

  \hline
\end{tabular}
\end{center}
Here we are working in units where the curvature radius of AdS$_5$ is $L=1$ and so we have $\alpha'^{-2}=4 \pi g_s N_c= g^2_{YM} N_c = \lambda$. The 10d Newton's and 5d Newton's constant are given by
\be (16 \pi G_{N,10})^{-1}=\frac{4 N^2}{(2 \pi)^5}, \quad (16 \pi G_N)^{-1} = V^S_5 (16 \pi G_{N,10})^{-1}
= \frac{N^2}{8 \pi^2}\ee
respectively, where $V^S_m$ is the volume of the unit $m$-sphere. The calculation of the EE in the D3/D5 and D3/D7 system now simply boils down to the determination of the EE in the two toy examples of the previous section with these particular values for the tension.

%%%%%%%%%%%%%%%%%%%%%%%%%%%%%%%%%
\section{Taming the UV divergences and Performing the Integrals}
\label{sec:doingintegrals}

\subsection{Preliminaries}

We derived a general prescription for calculating EEs in generic probe brane systems, eq.\eqref{doubleintegral}.
As a confirmation, we now calculate the EE using our formalism in the two solvable toy models
(the spacetime filling brane and the codimension-1 RS brane) introduced in section \ref{sec:toymodels}.
We'll compare our answer from the double integral, that is correct to leading order in the backreaction, to the exact expression,
which is possible to obtain in these two models as the fully backreacted metric is available in closed form.
We also showed in the last section that the leading order answers we calculate in these toy models are,
in fact, directly applicable to two of the most interesting top-down systems:
the D3/D7 system with the D7 wrapping AdS$_5$ $\times$ $S^3$ and the D3/D5 system with the D5 wrapping AdS$_4$ $\times$ $S^2$. For the special case of a spherical entangling surface we also compare our answers to an alternate method based on the trick of \cite{andykristan,Casini:2011kv}.

One subtlety we have to deal with is that the double integral eq.\eqref{doubleintegral} as it stands is naively doubly UV divergent:
both the $z$-integral over the probe brane worldvolume as well as the $w$-integral over the minimal area appear UV divergent.
We'll deal with these two UV divergences in turn and will calculate the integrals in our two examples.
We'll see that the UV divergence in the $z$-integral is a gauge artifact and can easily be removed.
The easiest way to do so is to evaluate the integrals by the not-even-trying method \cite{D'Hoker:1999ni}.
The UV-divergences in the $w$-integral on the other hand are physical, and reflect the fact that most contributions to the EE
are sensitive to short distance physics.
We explicitly regulate the double integral and evaluate all terms in the EE,
universal and cut-off dependent ones. In order to isolate the contributions
due to the probe brane, we subtract off the EE of the field theory without the probe.

\subsection{The probe brane $z$ integral}

While the double integral for the EE is symmetric between probe and minimal area, from the physical point of view it is most natural to first perform the $z$ integral to obtain the backreacted metric, and then to do the $w$ integral next in order to evaluate the resulting change in the EE. To do so, we need to first specify the propagator: both of our examples calculate the probe EE in a background AdS space; furthermore, the examples we consider all involve both a static probe brane and a static minimal area. So there is no time dependence involved and we can use the Euclidean AdS graviton propagator of \cite{D'Hoker:1999jc}:
\begin{eqnarray} \nonumber
 G_{\mu \nu \mu' \nu'}(w-z) &=& (\partial_{\mu} \partial_{\mu'} u \partial_{\nu} \partial_{\nu'} u +
  \partial_{\mu} \partial_{\nu'} u \partial_{\nu} \partial_{\mu'} u) G(u) + g_{\mu \nu} g_{\mu' \nu'} H(u)\\
  \nonumber
  &&+ (\partial_{(\mu} [ \partial_{\nu)} \partial_{\mu'} u \partial_{\nu'} u  X(u)] +
  (\partial_{(\mu'} [ \partial_{\nu')} \partial_{\mu} u \partial_{\nu} u  X(u)]\\
  \nonumber
  &&+ (\partial_{(\mu} [ \partial_{\nu)} u \partial_{\mu'} u \partial_{\nu'} u  Y(u)]+
  (\partial_{(\mu'} [ \partial_{\nu')} u \partial_{\mu} u \partial_{\nu} u  Y(u)]\\
  &&+ \partial_{\mu} [\partial_{\nu} Z(u)] g_{\mu'\nu'} +  \partial_{\mu'} [\partial_{\nu'} Z(u)] g_{\mu \nu}.
\end{eqnarray}
Here $u$ is the geodesic distance,
\be
u \equiv \frac{(z-w)^2}{2 z_0 w_0} = \frac{(\vec{z}-\vec{w})^2 +(z_0-w_0)^2}{2 z_0 w_0},
\ee
primed (unprimed) derivatives are derivatives with respect to $z$ ($w$), and
$(\cdots)$ denotes symmetrization with unit strength. The advantage of writing the propagator in this form is that it is manifestly engineered to isolate the gauge invariant information: the terms involving $X$, $Y$ and $Z$ are gradients with respect to $w$ or $z$. These $w$ gradient terms correspond to a gauge transformation we perform on the resulting $(\delta g)_{\mu \nu}(w)$ we get from integrating the propagator against $T^{probe}_{\mu' \nu'}(z)$. The $w$ gradients are annihilated by the differential operator ${\cal W}_{\mu \nu}^{\, \, \, \, \, \, \lambda \rho}$ from eq.\eqref{w} - Einstein's equations only determine the metric up to a gauge transformation. The gradients with respect to $z$ can be absorbed into the gauge transformation parameter $\Lambda_{\mu \nu \nu'}$ appearing on the right hand side of the defining differential equation for ${\cal W}_{\mu \nu}^{\, \, \, \, \, \, \lambda \rho}$, eq.\eqref{definegreen}. When contracted against conserved stress tensors, these total derivative terms can be integrated by parts and do not contribute as long as the sources vanish at infinity. The full gauge invariant information in the propagator is contained in the functions $G(u)$ and $H(u)$ which were determined in \cite{D'Hoker:1999jc}.

Unfortunately this beautiful formalism fails to give a finite answer for sources that extend to the boundary, as is the case for most probe brane systems of interest.  Close to the boundary, $u\rightarrow\infty$, $G$ and $H$ vanish as $G\sim u^{-d}$ and $H\sim u^{2-d}$. Taking into account the tensor structure in the propagator, for both terms the leading behavior near the boundary is $u^{4-d}$; taking the measure into account the integrand goes as $u^4$. Hence the source $T^{\mu \nu}_{probe}$ has to vanish faster than $u^4$ for the integral to remain finite. For a $T^{\mu \nu}_{probe}$ corresponding to a finite energy density this is the case. For a probe brane extending to infinity, where the non-vanishing components of $T^{\mu \nu}_{probe}$ typically go as $g^{\mu \nu} \sim u^{-2}$, the integral diverges as $u^2$. Since $H$ and $G$ multiply different tensor structures these divergences do not cancel against each other. What keeps the integral finite in the end are the ``gauge variant" terms in the propagator, $X$, $Y$ and $Z$. When integrated against the non-vanishing $T^{\mu \nu}_{probe}$ they leave divergent boundary terms behind which have to cancel the divergences due to $G$ and $H$. Unfortunately this means one needs to know the full propagator and not just $G$ and $H$.

Fortunately the same authors in \cite{D'Hoker:1999ni} put forward an alternative method to do $z$-integrals without-even-trying which automatically cures these UV divergences. Instead of calculating the integral directly, one applies ${\cal W}_{\mu \nu}^{\, \, \, \, \, \, \lambda \rho}$ on the integral
\be (\delta g)_{\mu \nu} =\int (d^{n+1} z \sqrt{g_I}) G_{\mu \nu \mu' \nu'} T^{\mu' \nu'}_{probe} \ee
and, using eq.\eqref{definegreen}, derives a simple differential equation for the integral itself. The astute reader will notice that this method essentially boils down to abandoning the Green's function approach and simply solving linearized Einstein's equations directly. There are however still advantages in thinking about the integral. For highly symmetric situations, as the one considered here, the inversion properties of the integral can be used to argue that the resulting metric deformation can only depend on a single variable, rendering Einstein's equations to be linear ODEs. Equivalence between the not-even-trying method with the direct evaluation of the integral was explicitly confirmed in \cite{D'Hoker:1999ni}. So, to do the integral, we can study the simpler problem of solving linearized Einstein's equations directly. The upshot is that the result of the $z$ integral are the metric perturbations eq.\eqref{codimzero} and eq.\eqref{codimone} for the codimension-0 and codimension-1 branes in AdS$_{d+1}$ respectively.

To be a little bit more precise let us quickly work through the codimension-1 case. Let $w_0$ denote the radial direction as before, $w_1$ the spatial direction orthogonal to the defect and ${\bf w}$ the spatial directions along the defect. Rotation and translation invariance in ${\bf w}$ directions tell us the $\delta {g}$ can not depend on ${\bf w}$ and that the only allowed tensor structures in
$\delta g$ are
\be
(\delta g)_{\mu \nu} = f_1 \,  g_{\mu \nu} + f_2  \, (P_0)_{\mu} (P_0)_{\nu} + f_3 \, (P_1)_{\mu} (P_1)_{\nu} + f_4 \, [ (P_1)_{\mu} (P_0)_{\mu} + (P_0)_{\nu} (P_1)_{\mu} ].
\ee
Here $(P_{0/1})_{\mu}$ are simply the projectors $\delta_{\mu}^{0/1}/w_0$.
In principle the functions $f_1$ through $f_4$ can depend on $w_0$ and $w_1$, but by invariance of the integral under scaling of $w_0$ and $w_1$, we can see that in fact they can only depend on $t=w_0/w_1$. Acting on $(\delta g)_{\mu \nu}$ with ${\cal W}_{\mu \nu}^{\, \, \, \, \, \, \lambda \rho}$ one derives a simple set of ODEs for $f_1$ through $f_4$ as a function of $t$ with the probe brane stress tensor evaluated at $w$ appearing as the source. Naively this set of differential equations looks over-determined, but it is easy to confirm that our $\delta g$ of eq.\eqref{codimone} indeed satisfies all of them as expected. The same method should still work if we include extra terms in the brane stress tensor, for example an electric field in the radial direction.

\subsection{The minimal area $w$ integral}

The $w$ integral is also UV divergent. Unlike the UV divergences in the $z$ integral, the UV divergencies in the $w$ integral are physical and reflect the short distance sensitivities of the EE. As long as the entangling surface intersects with the conformal defect,
there exists a UV divergence in the probe contribution to the EE due to the entanglement from those short-distance degrees of freedom of the defect theory.
They manifest themselves via the standard EE divergent structure of the conformal defect-localized sector. So in order to do the $w$ integral and to isolate and calculate these UV sensitive terms as well as the universal remainders, we need to chose a regularization procedure.
This turns out to be an essential but subtle step.

In order to isolate the EE due to the probe, we need to ensure that we do not modify the contribution from the non-probe degrees of freedom, dual to bulk gravity. It turns out, following \cite{Azeyanagi:2007qj}, that we need to modify the holographic
renormalization procedure for the leading contribution to the EE in order to ensure that this is the case. Due to the backreaction of the probe brane, the induced metric on the original cutoff slice will be slightly altered, and so will the leading order EE associated with the non-probe degrees of freedom. In order to avoid this effect we need to chose a new cut-off surface, constructed so that the induced metric on the cutoff slice is the same in the perturbed metric with the new cutoff as it was in the unperturbed metric with the old cutoff.
Formally we can write
\be
\delta A =
\left(
A\left[ g' \right]_{\Sigma'}  - A \left[g \right]_{\Sigma}
\right)_{
%\mbox{~with~}
\gamma_{\Sigma'}[g'] = \gamma_{\Sigma}[g],
}
\ee
with $A[g]_{\Sigma}$ being the area of the minimal surface calculated with the given metric $g$ and cutoff slice $\Sigma$,
and $\gamma_{\Sigma}[g]$ being the induced metric of $g$ onto $\Sigma$.
In particular, the constraining relation above can be solved,
providing the explicit definition of the new cutoff slice, $\Sigma'=\Sigma'(g'; g,\Sigma)$,
constructed in such a way that the induced metric for the boundary is the same for both $g$ on $\Sigma$ and $g'$ on $\Sigma'$.
Therefore, the backreaction from the probe-brane perturbation enters the final result not only as
the change of the integrand through metric perturbation, but also as the change of integration domain
from a new UV cut-off slice associated with the  perturbation.
In our formal expressions for the variation of the area as response to a metric perturbation, this effect was not visible, as it is all about regulating the UV divergences.

While in general this change of UV cutoff seems to be rather complicated, in the probe brane limit we are working with,
these two contributions disentangle nicely from each other to the leading order of probe brane tension.
We can write such disentangling effect formally as:
\be
\label{cutoffchange}
\delta A =
\left(A\left[ g+\delta g \right]_{\Sigma}  - A \left[g \right]_{\Sigma}\right)
+
\left(
 A \left[g \right]_{\Sigma'(g';g,\Sigma)}- A \left[g \right]_{\Sigma}
\right)
+ O\left(t_0^2 \right).
\ee
That is,
to correctly capture the effects to leading order in $t_0$,
we need to calculate the change in area due to the perturbed metric with the old cutoff,
as well as the change in the original area due to the new cutoff.
We can neglect the contribution from integrating the change in the area from the old to the new cutoff, as that term is order $t_0^2$.

The first regularized term allows the canonical treatment of the variational
principle with Dirichlet boundary condition, and our double integral formula applies straightforwardly.
The second subtraction term is constructed according to the principle of using the same holographic renormalization scheme
before and after perturbation. It admits a simple geometrical interpretation as the difference of the minimal surface bounded between
the original cutoff plane and the associated new cutoff plane.
In what follows we will see that it is crucial to include this extra contribution in order to get sensible answers for the probe EE.

\subsubsection{The codimension-1 RS brane and the D3/D5 system}

As a first consistency check of our procedure, we want to calculate the EE for the codimension-1 RS brane and compare it to the exact answer obtained from the fully backreacted metric. In addition, for this case of a conformal probe brane in a conformal background, an alternative method exists to get the EE. As recently pointed out by Jensen and O'Bannon, in this case one can apply the trick of \cite{Casini:2011kv} to determine the EE by mapping it to a thermal entropy in a hyperbolic space. We review the Jensen-O'Bannon calculation in the appendix. Reassuringly, we will find that all three calculations agree.

In order to apply our tools for the probe calculation, we first need to determine the correct UV cutoff on the $w$ integral.
For the EE calculation in the unperturbed $AdS_{d+1}$ background one typically chooses the flat cutoff plane $w_0^*=\epsilon$
in the AdS Poincar\'e coordinates.
To calculate the associated cutoff surface due to the perturbation in the toy model,
it is much easier to work with the perturbation in the original AdS-sliced coordinates as in eq.\eqref{ourmetric}:
\be
ds^2=dr^2+\frac{\cosh^2(|r|-c)}{y^2}(dy^2+dt^2+d\vec{w}^2)  \rightarrow
dr^2+\frac{\cosh^2r- 2 c \cosh r\mbox{~} \sinh |r|}{y^2}(dy^2+dt^2+d\vec{w}^2),
\ee
with AdS-radius $L$ scaled to 1.
In this coordinate system the original UV cut-off plane is located at $w_0^*=\frac{y}{\cosh r}=\epsilon$.
The perturbation-corrected cutoff surface is given by the following locus:
\be
\cosh r = \frac{y}{\epsilon} (1+c \frac{\sqrt{y^2 - \epsilon^2}}{y}).
\ee
With this subtraction scheme we present two cases for this toy model:
a spherical entangling surface bisected by the defect, and a strip entangling surface containing the defect.

\vskip10pt

{\bf Case 1 - Spherical Entangling Surface:} Choosing the spherical entangling surface bisected by the defect, the minimal surface in $AdS_{d+1}$ turns out to be \cite{Ryu:2006bv} the $d-1$ dimensional hemisphere
\be
R^2 = w_0^2 + w_1^2 +\vec{w}^2 = y^2 + \vec{w}^2.
\ee
Using the AdS-sliced coordinate, the induced metric on the minimal area and its stress tensor are given by:
\be
\sqrt{\gamma}=\left(\frac{\cosh r}{y}\right)^{d-2} (1-y^2)^{\frac{d-4}{2}} \, vol^S_{d-3},
\ee
\be
\frac{1}{2} \gamma^{ab} x^{\mu}_{,a}x^{\nu}_{,b}\delta g_{\mu\nu}= -c \tanh r * Tr[1\!\!1_{(d-2)}] = -c \, (d-2)\tanh r,
\ee
where $vol^S_{m}$ denotes the volume form of the unit $m$-sphere.

To calculate the EE we need to calculate the change in area from our double integral formula eq.\eqref{doubleintegral}, and then
subtract from it the contribution from the change of cutoff as indicated in eq.\eqref{cutoffchange}. Using the metric perturbation eq.\eqref{ourmetricperturbed} we have from the double integral (using $c_r\equiv \cosh r$ as an integration variable)
\be I_1 = 
\frac{1}{4G_N} V^S_{d-3} \, 
2 \, 
\int_{\epsilon}^1 dy \, \int_{1}^{\frac{y}{\epsilon}} dc_r \, [-c (d-2) c_r^{d-3}]
\frac{(1-y^2)^{\frac{d-4}{2}}}{y^{d-2}}. 
\ee
The factor of 2 comes about since we need to integrate from the midpoint ($c_r=1$) both to positive and negative infinity in $r$. By symmetry this is twice the integral to large positive $r$.
The original minimal area integrated from old and new cutoff gives us
\be
I_2 = 
\frac{1}{4 G_N} V^S_{d-3} \, 
2 \, 
\int_{\epsilon}^1 dy \, 
\int_{\frac{y}{\epsilon}}^{\frac{y}{\epsilon} (1 + c \sqrt{1-\frac{\epsilon^2}{y^2}})} dc_r \,  \frac{c_r^{d-2}}{\sqrt{c_r^2-1}}
 \frac{(1-y^2)^{\frac{d-4}{2}}}{y^{d-2}}. 
\ee
Note that the $y$-integral is already the same in both $I_1$ and $I_2$, but the $c_r$ integrals differ. Looking at the sum of $I_1$ and $I_2$ we see that the $c_r$ integral we need to do is
\be \tilde{I} \equiv -c (d-2) \int_{1}^{\frac{y}{\epsilon}} dc_r \,  c_r^{d-3} + \int_{\frac{y}{\epsilon}}^{\frac{y}{\epsilon} (1 + c \sqrt{1-\frac{\epsilon^2}{y^2}})} dc_r \, \frac{c_r^{d-2}}{\sqrt{c_r^2-1}}. \ee
As we are only interested in the leading $c \sim t_0$ behavior, we can expand out the second term in a power series
in $c$,
\be
\int_{\frac{y}{\epsilon}}^{\frac{y}{\epsilon} (1 + c \sqrt{1-\frac{\epsilon^2}{y^2}})} dc_r \, \frac{c_r^{d-2}}{\sqrt{c_r^2-1}}
=c \left ( \frac{y}{\epsilon} \right )^{d-2} + {\cal O}(c^2),
\ee
so that we simply get $\tilde{I}= c + {\cal O}(c^2)$. The probe contribution to the EE then becomes
\be
\label{linearized}
S_A = \frac{1}{2G_N} V^S_{d-3} c \int_{\epsilon}^1 dy \, \frac{(1-y^2)^{\frac{d-4}{2}}}{y^{d-2}} =
\frac{2 \pi T_0}{d-1} V^H_{d-2}.
\ee
Here we used eq.\eqref{t0} and eq.\eqref{tzero} (with the radius of curvature set to 1) to re-express $c$ in terms of $T_0$ and recognized the $y$ integral as the (regulated) volume of the unit hyperboloid as defined in appendix A. As also reviewed there, $V^H_{d-2}$ is exactly the structure to expect from a conformal field theory is $d-1$ dimensions; the EE for the defect degrees of freedom has the same functional form as that of a CFT living on the defect.

This answer is in perfect agreement with the Jensen-O'Bannon result of Appendix A. It is also in perfect agreement with the answer we get in the fully backreacted solution as we will now show. In the fully backreacted solution, the minimal area is locally given by the same embedding, so the worldvolume area element is still given by $\cosh^{d-2} \bar{r}$  where $\bar{r}=r-c$ for $r>0$. To isolate the EE contribution due to the defect, we still need to subtract out the EE for the theory without the defect. Matching cutoff procedures requires that we truncate both integrals at the same value of the warpfactor, $\cosh \bar{r}_*=\cosh r_*$. Since the $r$ integral starts at 0, whereas the $\bar{r}$ integral starts at $-c$, the difference is given by a $\bar{c}_r$ integral between these finite values. As $y$ no longer appears as an integration boundary, the two integrals factorize and the defect EE is given by
\be
\label{full}
S_A = \frac{1}{2G_N} V^S_{d-3} \int_0^{-cosh c} d\bar{c}_r \, \frac{\bar{c}_r^{d-2}}{\sqrt{\bar{c}_r^2}-1} \, \int_{\epsilon}^1 dy \, \frac{(1-y^2)^{\frac{d-4}{2}}}{y^{d-2}} = \frac{1}{2G_N} \, F(c) \, V^H_{d-2}.
\ee
Here
\be F(c) = \int_0^{\cosh c } dc_r \frac{c_r^{d-2}}{\sqrt{1-c_r^2}} \ee
is a non-linear function of $c=\tanh^{-1}(4 \pi G_N T_0/(d-1))$ that can easily be evaluated in terms of hypergeometric functions. What is important here is that, in the limit of small $c$, we have
\be F(c) = c + {\cal O} (c^2)  \ee
which can e.g. easily be seen by changing variables from $c_r$ to $x$, where $c_r=1+x^2/2$ and expanding the integral for small $x$. Plugging in $c\sim 4 \pi G_N T_0/(d-1)$ for small $c$, we see that the fully backreacted EE of eq.\eqref{full} indeed reduces to the probe limit answer of eq.\eqref{linearized} when linearizing the full answer in the brane tension.

\vskip10pt

\noindent {\bf Case 2 - Strip Entangling Surface:}
Choosing the static gauge for strip entangling surface  along the $x^1$-direction
in the Poincar\'e patch as eq.\eqref{adsmetric},
$  U^{d-1}=\{(w^0,\vec{w}) \} \hookrightarrow  AdS_{d+1}$,
one can verify that the minimal surface is given by solving the following equation:
$x^{d}$ being any constant (the specific time for the entangling surface),
$x^0=w^0$, $x^{i+1}=w^i$ for $i\in\{1,2,\ldots,d-2\}$, and
\be
\frac{d x^1}{dw^0}= \frac{\pm 1}{\sqrt{\left(\frac{L_U}{w^0}\right)^{2d-2}-1}}.
\ee
Using translational and scaling transformations, we can place the strip
in the following 2-branched parametrization located around the origin,
\be
\label{ushapesurface}
x^1=\pm L_U \left(
\frac{1}{d} \left(\frac{w^0 }{L_U }\right)^{d}
~ {}_2F_1
\left[
\frac{1}{2},
\frac{d}{2d-2},
\frac{3d-2}{2d-2},
\left(\frac{w^0}{L_U}\right)^{2d-2}
\right]
\mp
\frac{\sqrt{\pi }}{d} \frac{\Gamma\left[ \frac{3d-2}{2d-2} \right]}{\Gamma\left[\frac{2d-1}{2d-2}\right]}
\right),
\ee
with  $L_U$ being its endpoint in the radial $x^0$-coordinate,
$\mp \frac{L_U \sqrt{\pi }}{d} \frac{\Gamma\left[ \frac{3d-2}{2d-2} \right]}{\Gamma\left[\frac{2d-1}{2d-2}\right]}$
being its two endpoints in the perpendicular $x^1$-direction to the defect on the boundary.
For applying our formula eq.\eqref{doubleintegral}, we first notice that, 
\be
\label{nottouse1}
\sqrt{\gamma}=\frac{L^{d-1}}{(w^0)^{d-1} \sqrt{1-\left(\frac{w^0}{L_U}\right)^{2d-2}}}
\rightarrow \frac{1}{(w^0)^{d-2} \sqrt{(w^0)^2- (w^0)^{2d}}},
\ee
where we temporarily restore the factor of $AdS$-radius $L$,
given another length scale, strip-depth $L_U$, is also present, 
in order to correctly identify the scaling forms of both factors. With this,
we can then safely rescale the $w^0$-coordinate from $0$ to $1$ for the further evaluation.
The rest of integrand, $\gamma^{ab}x^{\mu}_{,a}x^{\nu}_{,b}\delta g_{\mu\nu}$ (the $\frac{1}{2}$ is
canceled due to the identical contribution from the two branches),
can be obtained by noticing that the metric perturbation eq.\eqref{codimone} separates
into two independent sectors of $x^{(0,1)}$-subspace and  $\vec{x}$-subspace:
For the  $x^{(0,1)}$-subspace,  given the metric perturbation is also in the form of direct product,
$\delta g_{\mu\nu} = \frac{-2\tilde{c} |x^1|}{(x^0)^2 \sqrt{(x^0)^2+(x^1)^2}} u_\mu u_\nu$ with
$u_\mu=\frac{(x^0,x^1)}{\sqrt{(x^0)^2+(x^1)^2}} $ and $\tilde{c}= \frac{t_0}{4(d-1)}$, we immediately have:
\be
\left(
\gamma^{ab}x^{\mu}_{,a}x^{\nu}_{,b}\delta g_{\mu\nu}
\right)_{\mbox{$x^{(0,1)}$-subspace}}=
\gamma^{00} \frac{-2\tilde{c} |x^1|}{ (x^0)^2 \sqrt{(x^0)^2+(x^1)^2}} (x^\mu_{,0}u_\mu) (x^\nu_{,0}u_\nu),
\ee
with
\be
\gamma_{00}=\frac{1+\left((x^1)'\right)^2}{(w^0)^2} \rightarrow
\gamma^{00}=(w^0)^2\left(1-(w^0)^{2d-2}\right),
\ee
and
\be
x^\mu_{,0}u_\mu
%=
%\left(\begin{array}{c}1 \\  \frac{1}{\sqrt{\left(\frac{L_U}{w^0}\right)^{2d-2}-1}} \end{array}\right)*
%\left(\begin{array}{cc}\frac{x^0}{(x^0)^2+(x^1)^2} & \frac{x^1}{(x^0)^2+(x^1)^2}   \end{array}\right)
\rightarrow
\frac{1}{(w^0)^2+(x^1)^2} \left( w^0 + \frac{x^1 (w^0)^d}{\sqrt{(w^0)^2 - (w^0)^{2d}}  }\right).
\ee
Consequently, we have
\be
\label{nottouse2}
\left(
\gamma^{ab}x^{\mu}_{,a}x^{\nu}_{,b}\delta g_{\mu\nu}
\right)_{\mbox{$x^{(0,1)}$-subspace}}
\rightarrow
 \frac{-2\tilde{c} |x^1|}{ \sqrt{(w^0)^2+(x^1)^2}}
\frac{
\left(1-(w^0)^{2d-2}\right)
}{(w^0)^2+(x^1)^2} \left( w^0 + \frac{x^1 (w^0)^d}{\sqrt{(w^0)^2 - (w^0)^{2d}}  }\right)^2.
\ee
For the $\vec{x}$-subspace, given the metric perturbation eq.\eqref{codimone} equals to the original $AdS_{d+1}$ metric
with an extra scaling factor $\frac{-2\tilde{c} |x^1|}{\sqrt{(x^0)^2+(x^1)^2}}$,
the result is
\be
\label{nottouse3}
\left(
\gamma^{ab}x^{\mu}_{,a}x^{\nu}_{,b}\delta g_{\mu\nu}
\right)_{\vec{x}\mbox{-subspace}} \rightarrow
\frac{-2\tilde{c} |x^1|}{\sqrt{(x^0)^2+(x^1)^2}} \mbox{Tr}[1\!\!1_{d-2}]= \frac{-2\tilde{c} |x^1|}{\sqrt{(x^0)^2+(x^1)^2}}(d-2).
\ee
Given eq.\eqref{nottouse1},\eqref{nottouse2},\eqref{nottouse3}, the final result for the strip entanglement entropy density,
after we factor out the translational invariant subspace spanned by $\vec{w}$,
with the constant cutoff slice $w^0=\epsilon$, is given by
%\be
%\label{ushaped1st}
%\int^{1}_{\epsilon}  \frac{dw^0}{(w^0)^{d-2} \sqrt{(w^0)^2-(w^0)^{2d}}} \frac{-2 \tilde{c} |x^1|}{\sqrt{(x^1)^2+(w^0)^2}}
%\left( d-2 + \frac{(w^0)^2-(w^0)^{2d}}{(x^1)^2+(w^0)^2} \left(1+\frac{x^1 (w^0)^{d-1}}{\sqrt{(w^0)^2-(w^0)^{2d}}}\right)^2 \right),
%\ee
\be
\label{ushaped1st}
\int^{1}_{\epsilon}
\frac{
dw^0 }{(w^0)^{d-2} \sqrt{(w^0)^2-(w^0)^{2d}}}
\frac{
(-2 \tilde{c} |x^1|)
\left( d-2 + \frac{(w^0)^2-(w^0)^{2d}}{(x^1)^2+(w^0)^2} \left(1+\frac{x^1 (w^0)^{d-1}}{\sqrt{(w^0)^2-(w^0)^{2d}}}\right)^2 \right)
}{\sqrt{(x^1)^2+(w^0)^2}}
,
\ee
together with the cutoff effect, eq.\eqref{cutoffchange},
which after some algebra can be rewritten succinctly into the following compact subtraction term:
\be
\label{ushaped2nd}
\frac{S_{sub}}{\mbox{Vol}^{d-2}_{Span\{\vec{w}\}}}
=
\left(
\frac{2\tilde{c} x^1}{\sqrt{(x^1)^2+(w^0)^2}}\frac{1}{(w^0)^{d-2}}
\right)_{w^0 \rightarrow \epsilon},
\ee
with
$\mbox{Vol}^{d-2}_{Span\{\vec{w}\}}$ being the volume of translational invariant $d-2$ dimensional subspace spanned by $\vec{w}$,
$x^1$ being the minimal embedding function eq.\eqref{ushapesurface} rescaled by $L_U \rightarrow 1$.
The change of minimal surface area density, according to eq.\eqref{cutoffchange},
will be given by the difference of eq.\eqref{ushaped1st} and eq.\eqref{ushaped2nd}.

Even though the above expressions seem daunting, simple numerical investigation nevertheless
reveals that the difference numerically evaluates to  $2.0 c$ for $AdS_4$ up to $AdS_7$,
before numerical instabilities render the evaluation inconclusive. After restoring the relevant scales,
this change of entanglement entropy density can be expressed as:
\be
s_{A}
=\frac{S_A}{\mbox{Vol}^{d-2}_{Span\{\vec{w}\}}}
= \frac{1}{4G_N} \left(2.0\ \tilde{c}\ \frac{L^{d-1}}{L_U^{d-2}}\right)
= \frac{2.0 \pi T_0}{(d-1)} \frac{L^{d}}{L_U^{d-2}},
\ \mbox{for $d \in \{4,5,6,7\}$},
\ee
with
$\tilde{c}$ being  re-expressed in terms of $T_0$ using eq.\eqref{t0} and eq.\eqref{tzero}.
Notice that this constant result also coincides with the result for $AdS_3$ obtained by \cite{Azeyanagi:2007qj}.
It is very tempting to conjecture that this numerical result also holds for any higher dimension. It should be possible to verify this simple answer analytically, but we will not pursue it further.

Note that in this case the answer for the defect contribution to the EE was UV finite. This is not too surprising, as the entangling surface did not intersect the defect and so there is no short range entanglement.

\subsubsection{The spacetime filling probe and the D3/D7 system}

Let us next calculate the EE for the spacetime filling probe brane, using our double integral prescription and once more compare it both to the exact answer from the fully backreacted metric as well as the Jensen-O'Bannon analysis from the appendix. The EE in this case is straight forward to calculate, as the bulk spacetime remains AdS$_{d+1}$ just with a shifted curvature radius given by eq.\eqref{lshift}. In order to see the consequences of this shift in $L$, let us restore the curvature radius in the expression for the EE for a spherical entangling surface of radius $R$ using an AdS$_{d+1}$ with curvature radius $L$ \cite{Ryu:2006ef}:
\be
S_A =\frac{L^{d-1}}{4 G_N} V^S_{d-2} \int_{R/a}^1 dy \, \frac{(1-y^2)^{(d-3)/2}}{y^{d-1}} =
\frac{L^{d-1}}{4 G_N}  V^H_{d-1}.
\ee
We see that the sphere radius only enters into the definition of $\epsilon$ in the regulated $V^H$. The overall $L^{d-1}$ factor cancels the dimensions in Newton's constant. Simply replacing $L$ with $l$ according to eq.\eqref{lshift} we see that the fully
backreacted contribution to the EE by the spacetime filling brane is (setting $L=1$ in the final answer to compare with expressions elsewhere in this paper)
\be S_A= \frac{l^{d-1} - L^{d-1}}{4 G_N} V^H_{d-1} = \frac{2 \pi T_0}{d}  V^H_{d-1}  + {\cal O}(t_0^2) \ee
in complete agreement with the Jensen-O'Bannon formula from the appendix.

To reproduce this calculation from our double integral formula, we once more need to add up two contributions. For one we have the direct contribution from the double integral, regulated by the old cutoff at\footnote{We require the short distance cut-off length $a=L \epsilon$ to have units of length for dimensionless $\epsilon$, this forces the factor of $L$ into this formula. Also, $z/L$ is the correct warpfactor.} $z/L=\epsilon$. In addition, we argued above that one should include a contribution from the changed cutoff, which is the integral of the original action from the old to the new cutoff. Matching cutoffs requires that $L/\epsilon=l/\epsilon$, so that the short distance cutoff $a$, which actually sets the range of our integrals, remains unchanged. Therefore, in the case of the codimension-0 brane, this second contribution vanishes.  Since $\delta g$ is AdS$_{d+1}$ with curvature radius $(\delta L)^2=t_0 L^2/[d (d-1)]$,
we have $T^{\mu \nu}_{min} \delta g_{\mu \nu}/2 = (d-1)/2$ and the double integral is in form identical to the original minimal area integral, just with a different prefactor and we obtain (again setting $L=1$ in the final answer)
\be
S_A = \frac{V^H_{d-1}}{4G_N} \frac{t_0 }{2d} = 2 \pi T_0 V^H_{d-1},
\ee
in perfect agreement with both the full non-linear formula and the Jensen-O'Bannon result.

%%%%%%%%%%%%%%%%%%%%%%%%%%%%%%%%
\section{Discussion of Results}
\label{sec:discussion}

In this work we determined the EE for a generic probe brane system as a formal double integral by explicitly taking into account the leading order backreaction of the brane. We validated our formula by working out the EE in various examples, comparing to established answers where possible.

It is somewhat surprising that in order to calculate the EE for a probe brane we actually seem to need to calculate at least the leading order backreaction. This has to be contrasted with calculations of the thermal entropy density, for which knowledge of the on-shell action of the probe itself is completely sufficient. As explained e.g. in \cite{Karch:2008uy} the thermal entropy can be obtained using thermodynamic identities from the free energy.
Latter can be obtained from the probe action alone. The bulk equations of motion ensure that the leading order backreaction, which naively comes in at the same order as the probe on-shell action, in fact does not contribute to the on-shell action. A direct calculation of the thermal entropy would also involve calculating the backreaction of the probe. Very similar to our calculation here, one would have to calculate the change in the horizon area due to the backreaction. In fact, as horizons are minimal area surfaces in the Euclidean black hole geometry, our formula directly applies to this case as well. As the horizon does typically not extend to the boundary, we even avoid the issues of UV divergences in the $w$-integral in that case.

By analogy one may hope that, at least in some circumstances, our double integral from eq.\eqref{doubleintegral} should simplify to an expression that localizes on the probe worldvolume alone. This is known to be true for the special case of a spherical entangling surfaces for conformal defects, as we made use of at various points in this work. There the calculation directly maps to a thermal entropy calculation and so it can once more be obtained via the free energy. The simple answer we get for the EE of a strip suggests that this, too, should follow from an easier calculation. The recent derivation of the RT prescription of \cite{Lewkowycz:2013nqa} hints that such a simplification should be possible more generally at least in static backgrounds. It would be interesting to make this more precise.

There are many potential future applications of our method. In this work we only looked at the simplest and most symmetric examples, both for the entangling surface and the probe embedding. More general entangling surfaces are straight forward. Recently the general terms appearing in the EE for boundary and hence also defect CFTs has been worked out (for $d=4$) in \cite{Fursaev:2013mxa}. Our results are in complete agreement with the structure found there, but trivially so. It would be interesting to confirm that the EE for more general entangling surfaces confirms the results of \cite{Fursaev:2013mxa}. Maybe more interestingly, our formula should be applied to some of the more interesting probe brane systems described in the introduction, where the EE can hopefully serve as a new probe to disentangle the interesting physics described by these probe branes. In particular, the finite density systems with a non-trivial worldvolume $F_{rt}$ should be easily within reach.

Our results can also be applied to gravitational theories including higher curvature corrections. In this case, it has been argued that the EE should be given by an appropriate Wald-like entropy \cite{Iyer:1994ys} associated with the minimal area \cite{Lewkowycz:2013nqa,Hung:2011xb}, even though it is not the standard Wald entropy. As long as the EE can be written as an integral of a local Lagrangian density over a bulk
surface, our formulism will still apply with $T^{min}_{\mu \nu}$ being the stress tensor associated with the new action. Of course higher derivative corrections are then also expected in the probe brane action, for which we allowed a general form already anyway.

\section*{Acknowledgments}
We would like to thank K.~Jensen and A.~O Bannon for useful discussions and especially for sharing their results prior to publication. AK would also like to thank the KMI at Nagoya University for hospitality during the final stages of this work. This work has been supported in part by the U.S. Department of Energy under Grant No.~DE-FG02-96ER40956.

\begin{appendix}
\section*{Appendix: The Jensen-O'Bannon Calculation}

Jensen and O'Bannon \cite{andykristan} pointed out that,
for the special case of spherical entangling surfaces
in a conformal theory where the probe preserves at least a lower dimensional subgroup of the conformal group (such as the D3/D5 and the D3/D7 system studied in the bulk of our paper at least in the absence of mass terms for the fundamental flavors), one can give an alternative derivation of the probe EE using the method of reference \cite{Casini:2011kv}. There it was pointed out that in a CFT the EE for a spherical entangling surface can be mapped to the thermal entropy associated with the same theory but formulated on a hyperboloid.
The radius of the sphere and the hyperboloid are equal, and as in a CFT this radius
sets the only scale, we will work with the unit sphere/unit hyperboloid. The
radius of the sphere can be restored by dimensional analysis in the end.
In the bulk this conformal transformation can easily be implemented by a change of coordinates. As long as we define the boundary metric by stripping off the quadratic divergence of the metric in the radial direction when approaching the boundary, different radial coordinates are associated with conformally related boundary metrics. AdS$_{d+1}$ can be written in a hyperbolic slicing as
\be
ds^2 = - h(r) dt^2 + \frac{dr^2}{h(r)} + r^2 dH^2_{d-1},
\ee
where $dH^2$ is the metric on the unit hyperboloid, which we can take to be given by
\be
\label{hyperboloid}
dH^2_{d-1} = d \rho^2 + \sinh^2 (\rho) \,  d \Omega^2_{d-2}.
\ee
AdS without a black hole corresponds to $h(r)=-1+r^2/L^2$. While this is the vacuum solution of Einstein's equations, in this coordinate system we see a horizon at $r=L$ with a temperature of $T=(2 \pi L)^{-1}$, see \cite{Vanzo:1997gw,Birmingham:1998nr,Emparan:1999pm,Emparan:1999gf}. The associated thermal entropy is the EE for the conformally related spherical entangling surface. The
thermal entropy of the hyperbolic horizon can be expressed as a finite
entropy density (which characterizes the number of degrees of freedom in
the dual field theory) times the volume of the hyperboloid. Latter is
of course infinite, but can be regulated by introducing a cutoff. In
terms of $y=(\cosh \rho)^{-1}$, we can define the volume of the $m$
dimensional
unit hyperboloid as the volume contained in the $y \geq \epsilon$ part
of the hyperboloid
\begin{eqnarray}
V^{H}_m &=& V^S_{m-1} \int_{\epsilon}^1 dy \, \frac{(1-y^2)^{(m-2)/2}}{y^{m}} \\
\nonumber
&=&  p_1 \left(\frac{1}{\epsilon}\right)^{m-1} + p_3 \left(\frac{1}{\epsilon}\right)^{m-3} + \ldots \\
&& \ldots + \left \{ \begin{array}{ll}
p_{m-1} \left(\frac{1}{\epsilon}\right) + p_{m} + {\cal O} (\epsilon) & m \mbox{ even} \\
p_{m-2} \left(\frac{1}{\epsilon}\right) + q \log(\epsilon) + {\cal O}(1) &  m \mbox{ odd} . \end{array}
\right .
\end{eqnarray}
Recall that $V^S_m$ denotes the volume of the unit $m$-sphere.
The coefficients $p_i$ can easily be determined from the integral expression.
They are spelled out explicitly in \cite{Ryu:2006ef}.

The free energy density of AdS in hyperbolic coordinates is
given by the gravitational on-shell action, which in turn simply
is the volume of space time
\be
\Omega = C_0 \int \sqrt{g} + {\cal L}_{ct} = C_0
\int_{r_h=L}^{\Lambda} r^{d-1} + {\cal L}_{ct},
\ee
with $C_0 = d/(8 \pi G_N)$.
As usual, this expression is divergent and can be regulated by counterterms as we indicated.
In fact, in order to systematically treat the $d$ dimensional case it was found to be easier to work with background subtraction, where here the correct background is the
zero temperature hyperbolic black hole, which corresponds to a black
hole with a negative mass parameter in hyperbolic slicing
\cite{Vanzo:1997gw,Birmingham:1998nr,Emparan:1999pm,Emparan:1999gf}.
In these papers it was explicitly confirmed that in low dimensions the regulation by counterterms is equivalent to using background subtraction.
The corresponding entropy density was found to be
(in units of the radius of the hyperboloid)
\be
\label{s}
s = \frac{2 \pi C_0}{d}.
\ee
As expected, this is just the standard Bekenstein-Hawking entropy, $1/(4G_N)$ times the horizon area.

As the EE for a spherical entangling
surface of unit radius is equal to this thermal entropy, it can easily be expressed in terms of $V^{H}_m$. Restoring the radius of the sphere, one finds for the EE associated with a sphere of radius $R$ \cite{Ryu:2006ef}:
\be
S = \frac{V^H_{d-1}}{4 G_N},
\ee
where the relation between $G_N$ and field theory quantities depends on the theory in question and the $\epsilon$'s appearing in $V^H$ should be read as $R/a$ where $a$ is the short distance cutoff length.

The probe branes of interest for us extend along a minimal AdS$_{n+1}$ slice inside AdS$_{d+1}$. In the original flat slicing, these minimal AdS$_{n+1}$ branes are obtained by setting some of the spatial components of $w_{\mu}$ to zero. In the hyperbolic coordinates, they are wrapping an equatorial $S^{n-2}$ inside the $S^{d-2}$ in
eq.\eqref{hyperboloid}. The worldvolume action is just the volume of the probe brane, so the free energy density is once more given by the regulated volume. For $n=d$ the calculation proceeds as above and we get
\be
S= \frac{2 \pi T_0}{d} V^H_{d-1}.
\ee
$T_0$ here is the effective tension, which for probe branes wrapping a product manifold is the full brane tension times the volume of the internal space, just as we found to be the case more generally in section \ref{sec:internal}.
For $n<d$ we also calculate the volume. But one should note that for the background subtraction we are still using the $d+1$ dimensional zero temperature hyperbolic black hole. It's easy to see that for $n<d$ background subtraction simply cancels the UV divergent terms without leaving a finite remainder and the free energy is given by $-T_0 r_h^{n-1}$. To get the entropy density from this, one needs the relation between $T$ and $r_h$, which for a hyperbolic black hole at $T=(2 \pi L)^{-1}$ is given by $r_H = 2 \pi T/(d-1)$, instead of the familiar $r_H = 4 \pi T/d$ from flat slicing black holes. With this we get for the entropy for $n<d$:
\be
S= \frac{2 \pi T_0}{d-1} V^H_{n-1}.
\ee
That is, the EE is equivalent to that of a spherical entangling surface in a $n+1$ dimensional field theory, with the degrees of freedom counted by $(2 \pi T_0)/(d-1)$ instead of $1/(4 G_N)$.

\end{appendix}

%%%%%%%%%%%%%%%%%
\bibliographystyle{JHEP}
\bibliography{probeEE}

\providecommand{\href}[2]{#2}\begingroup\raggedright\begin{thebibliography}{10}

\bibitem{Witten:1998qj}
E.~Witten, {\it {Anti-de Sitter space and holography}},  {\em
  Adv.Theor.Math.Phys.} {\bf 2} (1998) 253--291,
  [\href{http://xxx.lanl.gov/abs/hep-th/9802150}{{\tt hep-th/9802150}}].

\bibitem{Maldacena:1997re}
J.~M. Maldacena, {\it {The Large N limit of superconformal field theories and
  supergravity}},  {\em Adv.Theor.Math.Phys.} {\bf 2} (1998) 231--252,
  [\href{http://xxx.lanl.gov/abs/hep-th/9711200}{{\tt hep-th/9711200}}].

\bibitem{Gubser:1998bc}
S.~Gubser, I.~R. Klebanov, and A.~M. Polyakov, {\it {Gauge theory correlators
  from noncritical string theory}},  {\em Phys.Lett.} {\bf B428} (1998)
  105--114, [\href{http://xxx.lanl.gov/abs/hep-th/9802109}{{\tt
  hep-th/9802109}}].

\bibitem{Ryu:2006bv}
S.~Ryu and T.~Takayanagi, {\it {Holographic derivation of entanglement entropy
  from AdS/CFT}},  {\em Phys.Rev.Lett.} {\bf 96} (2006) 181602,
  [\href{http://xxx.lanl.gov/abs/hep-th/0603001}{{\tt hep-th/0603001}}].

\bibitem{Takayanagi:2012kg}
T.~Takayanagi, {\it {Entanglement Entropy from a Holographic Viewpoint}},  {\em
  Class.Quant.Grav.} {\bf 29} (2012) 153001,
  [\href{http://xxx.lanl.gov/abs/1204.2450}{{\tt arXiv:1204.2450}}].

\bibitem{Karch:2002sh}
A.~Karch and E.~Katz, {\it {Adding flavor to AdS / CFT}},  {\em JHEP} {\bf
  0206} (2002) 043, [\href{http://xxx.lanl.gov/abs/hep-th/0205236}{{\tt
  hep-th/0205236}}].

\bibitem{Karch:2000gx}
A.~Karch and L.~Randall, {\it {Open and closed string interpretation of SUSY
  CFT's on branes with boundaries}},  {\em JHEP} {\bf 0106} (2001) 063,
  [\href{http://xxx.lanl.gov/abs/hep-th/0105132}{{\tt hep-th/0105132}}].

\bibitem{Sakai:2004cn}
T.~Sakai and S.~Sugimoto, {\it {Low energy hadron physics in holographic QCD}},
   {\em Prog.Theor.Phys.} {\bf 113} (2005) 843--882,
  [\href{http://xxx.lanl.gov/abs/hep-th/0412141}{{\tt hep-th/0412141}}].

\bibitem{Karch:2007pd}
A.~Karch and A.~O'Bannon, {\it {Metallic AdS/CFT}},  {\em JHEP} {\bf 0709}
  (2007) 024, [\href{http://xxx.lanl.gov/abs/0705.3870}{{\tt
  arXiv:0705.3870}}].

\bibitem{Hartnoll:2009ns}
S.~A. Hartnoll, J.~Polchinski, E.~Silverstein, and D.~Tong, {\it {Towards
  strange metallic holography}},  {\em JHEP} {\bf 1004} (2010) 120,
  [\href{http://xxx.lanl.gov/abs/0912.1061}{{\tt arXiv:0912.1061}}].

\bibitem{Kachru:2009xf}
S.~Kachru, A.~Karch, and S.~Yaida, {\it {Holographic Lattices, Dimers, and
  Glasses}},  {\em Phys.Rev.} {\bf D81} (2010) 026007,
  [\href{http://xxx.lanl.gov/abs/0909.2639}{{\tt arXiv:0909.2639}}].

\bibitem{Jensen:2011su}
K.~Jensen, S.~Kachru, A.~Karch, J.~Polchinski, and E.~Silverstein, {\it
  {Towards a holographic marginal Fermi liquid}},  {\em Phys.Rev.} {\bf D84}
  (2011) 126002, [\href{http://xxx.lanl.gov/abs/1105.1772}{{\tt
  arXiv:1105.1772}}].

\bibitem{Jensen:2010ga}
K.~Jensen, A.~Karch, D.~T. Son, and E.~G. Thompson, {\it {Holographic
  Berezinskii-Kosterlitz-Thouless Transitions}},  {\em Phys.Rev.Lett.} {\bf
  105} (2010) 041601, [\href{http://xxx.lanl.gov/abs/1002.3159}{{\tt
  arXiv:1002.3159}}].

\bibitem{Karch:2008fa}
A.~Karch, D.~Son, and A.~Starinets, {\it {Zero Sound from Holography}},
  \href{http://xxx.lanl.gov/abs/0806.3796}{{\tt arXiv:0806.3796}}.

\bibitem{Kulaxizi:2008jx}
M.~Kulaxizi and A.~Parnachev, {\it {Holographic Responses of Fermion Matter}},
  {\em Nucl.Phys.} {\bf B815} (2009) 125--141,
  [\href{http://xxx.lanl.gov/abs/0811.2262}{{\tt arXiv:0811.2262}}].

\bibitem{Karch:2009eb}
A.~Karch, M.~Kulaxizi, and A.~Parnachev, {\it {Notes on Properties of
  Holographic Matter}},  {\em JHEP} {\bf 0911} (2009) 017,
  [\href{http://xxx.lanl.gov/abs/0908.3493}{{\tt arXiv:0908.3493}}].

\bibitem{Chang:2012ek}
H.-C. Chang and A.~Karch, {\it {Novel Solutions of Finite-Density D3/D5 Probe
  Brane System and Their Implications for Stability}},  {\em JHEP} {\bf 1210}
  (2012) 069, [\href{http://xxx.lanl.gov/abs/1207.7078}{{\tt
  arXiv:1207.7078}}].

\bibitem{Ammon:2012mu}
M.~Ammon, K.~Jensen, K.-Y. Kim, J.~N. Laia, and A.~O'Bannon, {\it {Moduli
  Spaces of Cold Holographic Matter}},  {\em JHEP} {\bf 1211} (2012) 055,
  [\href{http://xxx.lanl.gov/abs/1208.3197}{{\tt arXiv:1208.3197}}].

\bibitem{Ammon:2012je}
M.~Ammon, M.~Kaminski, and A.~Karch, {\it {Hyperscaling-Violation on Probe
  D-Branes}},  {\em JHEP} {\bf 1211} (2012) 028,
  [\href{http://xxx.lanl.gov/abs/1207.1726}{{\tt arXiv:1207.1726}}].

\bibitem{Ogawa:2011bz}
N.~Ogawa, T.~Takayanagi, and T.~Ugajin, {\it {Holographic Fermi Surfaces and
  Entanglement Entropy}},  {\em JHEP} {\bf 1201} (2012) 125,
  [\href{http://xxx.lanl.gov/abs/1111.1023}{{\tt arXiv:1111.1023}}].

\bibitem{Huijse:2011ef}
L.~Huijse, S.~Sachdev, and B.~Swingle, {\it {Hidden Fermi surfaces in
  compressible states of gauge-gravity duality}},  {\em Phys.Rev.} {\bf B85}
  (2012) 035121, [\href{http://xxx.lanl.gov/abs/1112.0573}{{\tt
  arXiv:1112.0573}}].

\bibitem{Fujita:2009kw}
M.~Fujita, W.~Li, S.~Ryu, and T.~Takayanagi, {\it {Fractional Quantum Hall
  Effect via Holography: Chern-Simons, Edge States, and Hierarchy}},  {\em
  JHEP} {\bf 0906} (2009) 066, [\href{http://xxx.lanl.gov/abs/0901.0924}{{\tt
  arXiv:0901.0924}}].

\bibitem{HoyosBadajoz:2010ac}
C.~Hoyos-Badajoz, K.~Jensen, and A.~Karch, {\it {A Holographic Fractional
  Topological Insulator}},  {\em Phys.Rev.} {\bf D82} (2010) 086001,
  [\href{http://xxx.lanl.gov/abs/1007.3253}{{\tt arXiv:1007.3253}}].

\bibitem{Azeyanagi:2007qj}
T.~Azeyanagi, A.~Karch, T.~Takayanagi, and E.~G. Thompson, {\it {Holographic
  calculation of boundary entropy}},  {\em JHEP} {\bf 0803} (2008) 054--054,
  [\href{http://xxx.lanl.gov/abs/0712.1850}{{\tt arXiv:0712.1850}}].

\bibitem{Chiodaroli:2010ur}
M.~Chiodaroli, M.~Gutperle, and L.-Y. Hung, {\it {Boundary entropy of
  supersymmetric Janus solutions}},  {\em JHEP} {\bf 1009} (2010) 082,
  [\href{http://xxx.lanl.gov/abs/1005.4433}{{\tt arXiv:1005.4433}}].

\bibitem{D'Hoker:1999jc}
E.~D'Hoker, D.~Z. Freedman, S.~D. Mathur, A.~Matusis, and L.~Rastelli, {\it
  {Graviton and gauge boson propagators in AdS(d+1)}},  {\em Nucl.Phys.} {\bf
  B562} (1999) 330--352, [\href{http://xxx.lanl.gov/abs/hep-th/9902042}{{\tt
  hep-th/9902042}}].

\bibitem{Hubeny:2007xt}
V.~E. Hubeny, M.~Rangamani, and T.~Takayanagi, {\it {A Covariant holographic
  entanglement entropy proposal}},  {\em JHEP} {\bf 0707} (2007) 062,
  [\href{http://xxx.lanl.gov/abs/0705.0016}{{\tt arXiv:0705.0016}}].

\bibitem{Nozaki:2013vta}
M.~Nozaki, T.~Numasawa, A.~Prudenziati, and T.~Takayanagi, {\it {Dynamics of
  Entanglement Entropy from Einstein Equation}},
  \href{http://xxx.lanl.gov/abs/1304.7100}{{\tt arXiv:1304.7100}}.

\bibitem{Wong:2013gua}
G.~Wong, I.~Klich, L.~A.~P. Zayas, and D.~Vaman, {\it {Entanglement Temperature
  and Entanglement Entropy of Excited States}},
  \href{http://xxx.lanl.gov/abs/1305.3291}{{\tt arXiv:1305.3291}}.

\bibitem{Hikida:2009tp}
Y.~Hikida, W.~Li, and T.~Takayanagi, {\it {ABJM with Flavors and FQHE}},  {\em
  JHEP} {\bf 0907} (2009) 065, [\href{http://xxx.lanl.gov/abs/0903.2194}{{\tt
  arXiv:0903.2194}}].

\bibitem{Gaiotto:2009tk}
D.~Gaiotto and D.~L. Jafferis, {\it {Notes on adding D6 branes wrapping RP**3
  in AdS(4) x CP**3}},  {\em JHEP} {\bf 1211} (2012) 015,
  [\href{http://xxx.lanl.gov/abs/0903.2175}{{\tt arXiv:0903.2175}}].

\bibitem{Hohenegger:2009as}
S.~Hohenegger and I.~Kirsch, {\it {A Note on the holography of Chern-Simons
  matter theories with flavour}},  {\em JHEP} {\bf 0904} (2009) 129,
  [\href{http://xxx.lanl.gov/abs/0903.1730}{{\tt arXiv:0903.1730}}].

\bibitem{Ammon:2009wc}
M.~Ammon, J.~Erdmenger, R.~Meyer, A.~O'Bannon, and T.~Wrase, {\it {Adding
  Flavor to AdS(4)/CFT(3)}},  {\em JHEP} {\bf 0911} (2009) 125,
  [\href{http://xxx.lanl.gov/abs/0909.3845}{{\tt arXiv:0909.3845}}].

\bibitem{Jensen:2010vx}
K.~Jensen, {\it {More Holographic Berezinskii-Kosterlitz-Thouless
  Transitions}},  {\em Phys.Rev.} {\bf D82} (2010) 046005,
  [\href{http://xxx.lanl.gov/abs/1006.3066}{{\tt arXiv:1006.3066}}].

\bibitem{Aharony:2008ug}
O.~Aharony, O.~Bergman, D.~L. Jafferis, and J.~Maldacena, {\it {N=6
  superconformal Chern-Simons-matter theories, M2-branes and their gravity
  duals}},  {\em JHEP} {\bf 0810} (2008) 091,
  [\href{http://xxx.lanl.gov/abs/0806.1218}{{\tt arXiv:0806.1218}}].

\bibitem{Karch:2008uy}
A.~Karch, A.~O'Bannon, and E.~Thompson, {\it {The Stress-Energy Tensor of
  Flavor Fields from AdS/CFT}},  {\em JHEP} {\bf 0904} (2009) 021,
  [\href{http://xxx.lanl.gov/abs/0812.3629}{{\tt arXiv:0812.3629}}].

\bibitem{Randall:1999ee}
L.~Randall and R.~Sundrum, {\it {A Large mass hierarchy from a small extra
  dimension}},  {\em Phys.Rev.Lett.} {\bf 83} (1999) 3370--3373,
  [\href{http://xxx.lanl.gov/abs/hep-ph/9905221}{{\tt hep-ph/9905221}}].

\bibitem{Randall:1999vf}
L.~Randall and R.~Sundrum, {\it {An Alternative to compactification}},  {\em
  Phys.Rev.Lett.} {\bf 83} (1999) 4690--4693,
  [\href{http://xxx.lanl.gov/abs/hep-th/9906064}{{\tt hep-th/9906064}}].

\bibitem{Israel:1966rt}
W.~Israel, {\it {Singular hypersurfaces and thin shells in general
  relativity}},  {\em Nuovo Cim.} {\bf B44S10} (1966) 1.

\bibitem{DeWolfe:2001pq}
O.~DeWolfe, D.~Z. Freedman, and H.~Ooguri, {\it {Holography and defect
  conformal field theories}},  {\em Phys.Rev.} {\bf D66} (2002) 025009,
  [\href{http://xxx.lanl.gov/abs/hep-th/0111135}{{\tt hep-th/0111135}}].

\bibitem{Kirsch:2005uy}
I.~Kirsch and D.~Vaman, {\it {The D3 / D7 background and flavor dependence of
  Regge trajectories}},  {\em Phys.Rev.} {\bf D72} (2005) 026007,
  [\href{http://xxx.lanl.gov/abs/hep-th/0505164}{{\tt hep-th/0505164}}].

\bibitem{D'Hoker:2007xy}
E.~D'Hoker, J.~Estes, and M.~Gutperle, {\it {Exact half-BPS Type IIB interface
  solutions. I. Local solution and supersymmetric Janus}},  {\em JHEP} {\bf
  0706} (2007) 021, [\href{http://xxx.lanl.gov/abs/0705.0022}{{\tt
  arXiv:0705.0022}}].

\bibitem{D'Hoker:2007xz}
E.~D'Hoker, J.~Estes, and M.~Gutperle, {\it {Exact half-BPS Type IIB interface
  solutions. II. Flux solutions and multi-Janus}},  {\em JHEP} {\bf 0706}
  (2007) 022, [\href{http://xxx.lanl.gov/abs/0705.0024}{{\tt
  arXiv:0705.0024}}].

\bibitem{Aharony:2011yc}
O.~Aharony, L.~Berdichevsky, M.~Berkooz, and I.~Shamir, {\it {Near-horizon
  solutions for D3-branes ending on 5-branes}},  {\em Phys.Rev.} {\bf D84}
  (2011) 126003, [\href{http://xxx.lanl.gov/abs/1106.1870}{{\tt
  arXiv:1106.1870}}].

\bibitem{Bigazzi:2005md}
F.~Bigazzi, R.~Casero, A.~Cotrone, E.~Kiritsis, and A.~Paredes, {\it
  {Non-critical holography and four-dimensional CFT's with fundamentals}},
  {\em JHEP} {\bf 0510} (2005) 012,
  [\href{http://xxx.lanl.gov/abs/hep-th/0505140}{{\tt hep-th/0505140}}].

\bibitem{Casero:2006pt}
R.~Casero, C.~Nunez, and A.~Paredes, {\it {Towards the string dual of N=1
  SQCD-like theories}},  {\em Phys.Rev.} {\bf D73} (2006) 086005,
  [\href{http://xxx.lanl.gov/abs/hep-th/0602027}{{\tt hep-th/0602027}}].

\bibitem{Mateos:2006yd}
D.~Mateos, R.~C. Myers, and R.~M. Thomson, {\it {Holographic viscosity of
  fundamental matter}},  {\em Phys.Rev.Lett.} {\bf 98} (2007) 101601,
  [\href{http://xxx.lanl.gov/abs/hep-th/0610184}{{\tt hep-th/0610184}}].

\bibitem{Bigazzi:2013jqa}
F.~Bigazzi, A.~L. Cotrone, and J.~Tarrio, {\it {Charged D3-D7 plasmas: novel
  solutions, extremality and stability issues}},  {\em JHEP} {\bf 1307} (2013)
  074, [\href{http://xxx.lanl.gov/abs/1304.4802}{{\tt arXiv:1304.4802}}].

\bibitem{Son:2002sd}
D.~T. Son and A.~O. Starinets, {\it {Minkowski space correlators in AdS / CFT
  correspondence: Recipe and applications}},  {\em JHEP} {\bf 0209} (2002) 042,
  [\href{http://xxx.lanl.gov/abs/hep-th/0205051}{{\tt hep-th/0205051}}].

\bibitem{Turyn:1988af}
M.~Turyn, {\it The graviton propagator in maximally symmetric spaces},  {\em
  J.Math.Phys.} {\bf 31} (1990) 669.

\bibitem{Allen:1986tt}
B.~Allen and M.~Turyn, {\it An evaluation of the graviton propagator in de
  {S}itter space},  {\em Nucl.Phys.} {\bf B292} (1987) 813.

\bibitem{Freund:1980xh}
P.~G. Freund and M.~A. Rubin, {\it {Dynamics of Dimensional Reduction}},  {\em
  Phys.Lett.} {\bf B97} (1980) 233--235.

\bibitem{DeWolfe:2001nz}
O.~DeWolfe, D.~Z. Freedman, S.~S. Gubser, G.~T. Horowitz, and I.~Mitra, {\it
  {Stability of AdS(p) x M(q) compactifications without supersymmetry}},  {\em
  Phys.Rev.} {\bf D65} (2002) 064033,
  [\href{http://xxx.lanl.gov/abs/hep-th/0105047}{{\tt hep-th/0105047}}].

\bibitem{andykristan}
K.~Jensen and A.~O'Bannon {\em {to appear}}.

\bibitem{Casini:2011kv}
H.~Casini, M.~Huerta, and R.~C. Myers, {\it {Towards a derivation of
  holographic entanglement entropy}},  {\em JHEP} {\bf 1105} (2011) 036,
  [\href{http://xxx.lanl.gov/abs/1102.0440}{{\tt arXiv:1102.0440}}].

\bibitem{D'Hoker:1999ni}
E.~D'Hoker, D.~Z. Freedman, and L.~Rastelli, {\it {AdS / CFT four point
  functions: How to succeed at z integrals without really trying}},  {\em
  Nucl.Phys.} {\bf B562} (1999) 395--411,
  [\href{http://xxx.lanl.gov/abs/hep-th/9905049}{{\tt hep-th/9905049}}].

\bibitem{Ryu:2006ef}
S.~Ryu and T.~Takayanagi, {\it {Aspects of Holographic Entanglement Entropy}},
  {\em JHEP} {\bf 0608} (2006) 045,
  [\href{http://xxx.lanl.gov/abs/hep-th/0605073}{{\tt hep-th/0605073}}].

\bibitem{Lewkowycz:2013nqa}
A.~Lewkowycz and J.~Maldacena, {\it {Generalized gravitational entropy}},
  \href{http://xxx.lanl.gov/abs/1304.4926}{{\tt arXiv:1304.4926}}.

\bibitem{Fursaev:2013mxa}
D.~Fursaev, {\it {Quantum Entanglement on Boundaries}},
  \href{http://xxx.lanl.gov/abs/1305.2334}{{\tt arXiv:1305.2334}}.

\bibitem{Iyer:1994ys}
V.~Iyer and R.~M. Wald, {\it {Some properties of Noether charge and a proposal
  for dynamical black hole entropy}},  {\em Phys.Rev.} {\bf D50} (1994)
  846--864, [\href{http://xxx.lanl.gov/abs/gr-qc/9403028}{{\tt
  gr-qc/9403028}}].

\bibitem{Hung:2011xb}
L.-Y. Hung, R.~C. Myers, and M.~Smolkin, {\it {On Holographic Entanglement
  Entropy and Higher Curvature Gravity}},  {\em JHEP} {\bf 1104} (2011) 025,
  [\href{http://xxx.lanl.gov/abs/1101.5813}{{\tt arXiv:1101.5813}}].

\bibitem{Vanzo:1997gw}
L.~Vanzo, {\it {Black holes with unusual topology}},  {\em Phys.Rev.} {\bf D56}
  (1997) 6475--6483, [\href{http://xxx.lanl.gov/abs/gr-qc/9705004}{{\tt
  gr-qc/9705004}}].

\bibitem{Birmingham:1998nr}
D.~Birmingham, {\it {Topological black holes in Anti-de Sitter space}},  {\em
  Class.Quant.Grav.} {\bf 16} (1999) 1197--1205,
  [\href{http://xxx.lanl.gov/abs/hep-th/9808032}{{\tt hep-th/9808032}}].

\bibitem{Emparan:1999pm}
R.~Emparan, C.~V. Johnson, and R.~C. Myers, {\it {Surface terms as counterterms
  in the AdS / CFT correspondence}},  {\em Phys.Rev.} {\bf D60} (1999) 104001,
  [\href{http://xxx.lanl.gov/abs/hep-th/9903238}{{\tt hep-th/9903238}}].

\bibitem{Emparan:1999gf}
R.~Emparan, {\it {AdS / CFT duals of topological black holes and the entropy of
  zero energy states}},  {\em JHEP} {\bf 9906} (1999) 036,
  [\href{http://xxx.lanl.gov/abs/hep-th/9906040}{{\tt hep-th/9906040}}].

\end{thebibliography}\endgroup

\end{document}